\documentclass[journal]{IEEEtran}

\usepackage{booktabs}
\usepackage{amsmath}
\usepackage{amsthm}
\usepackage{multirow}
\usepackage{color,cite}
\usepackage{subfigure}
\usepackage{mathrsfs}
\usepackage{amssymb}
\usepackage{bm}
\usepackage{graphicx}
\usepackage{amsfonts}
\usepackage{algorithm}
\usepackage{algorithmic}
\usepackage{epstopdf}

\newcommand{\tabincell}[2]{\begin{tabular}{@{}#1@{}}#2\end{tabular}}

\begin{document}

\bibliographystyle{IEEEtran}

\title{\Huge{Deep Learning Assisted Calibrated Beam Training \\ for Millimeter-Wave Communication Systems}}

\author{
\large{Ke Ma, \emph{Student Member, IEEE}, Dongxuan He, Hancun Sun, \\Zhaocheng Wang, \emph{Fellow, IEEE}, and Sheng Chen, \emph{Fellow, IEEE}}
\thanks{This work was supported in part by the National Key R\&D Program of China under Grant 2018YFB1801102 and in part by National Natural Science Foundation of China (Grant No. 61871253). \emph{(Corresponding author: Zhaocheng Wang, Dongxuan He.)}}
\thanks{K.~Ma, D. He, H. Sun and Z. Wang are with Beijing National Research Center for Information Science and Technology, Department of Electronic Engineering, Tsinghua University, Beijing 100084, China, and Z. Wang is also with Tsinghua Shenzhen International Graduate School, Shenzhen 518055, China (E-mails: ma-k19@mails.tsinghua.edu.cn, dongxuan\_he@mail.tsinghua.edu.cn, sunhc18@mails.tsinghua.edu.cn, zcwang@tsinghua.edu.cn).} %
\thanks{S. Chen is with the School of Electronics and Computer Science, University of Southampton, Southampton SO17 1BJ, U.K. (E-mail: sqc@ecs.soton.ac.uk).}
\vspace*{-6mm}
}

\maketitle

\begin{abstract}
Huge overhead of beam training imposes a significant challenge in millimeter-wave (mmWave) wireless communications. To address this issue, in this paper, we propose a wide beam based training approach to calibrate the narrow beam direction according to the channel power leakage. To handle the complex nonlinear properties of the channel power leakage, deep learning is utilized to predict the optimal narrow beam directly.
Specifically, three deep learning assisted calibrated beam training schemes are proposed. The first scheme  adopts convolution neural network to implement the prediction based on the instantaneous received signals of wide beam training. We also perform the additional narrow beam training based on the predicted probabilities for further beam direction calibrations. However, the first scheme only depends on one wide beam training, which lacks the robustness to noise. To tackle this problem, the second scheme adopts long-short term memory (LSTM) network for tracking the movement of users and calibrating the beam direction according to the received signals of prior beam training, in order to enhance the robustness to noise. To further reduce the overhead of wide beam training, our third scheme, an adaptive beam training strategy, selects partial wide beams to be trained based on the prior received signals. Two criteria, namely, optimal neighboring criterion and maximum probability criterion, are designed for the selection. Furthermore, to handle mobile scenarios, auxiliary LSTM is introduced to calibrate the directions of the selected wide beams more precisely. Simulation results demonstrate that our proposed schemes achieve significantly higher beamforming gain with smaller beam training overhead compared with the conventional and existing deep-learning based counterparts.
\end{abstract}


\IEEEpeerreviewmaketitle

\section{Introduction}\label{Sec:INTRO} 

Benefiting from huge bandwidth resources in the millimeter-wave (mmWave) frequency band, mmWave communication enables high transmission rates and it offers one of the crucial technologies in the fifth-generation (5G) wireless communication systems \cite{Ref:MMWAVE1,Ref:MMWAVE2}. However, mmWave signals suffer from more serious pathloss compared with the conventional sub-6GHz communication systems \cite{Ref:PATHLOSS}. Fortunately, the small wavelength of mmWave signals allows more antennas to be integrated into both base stations (BSs) and user equipment (UE) \cite{Ref:ANTENNA}. Therefore, large antenna arrays can be equipped at BS and UE sides to implement directional beamforming, such that high pathloss can be compensated by the beamforming gain \cite{Ref:BEAMFORMING1,Ref:BEAMFORMING2,Ref:BEAMFORMING3}.

In order to enhance the received power of mmWave signals, beam training has been widely adopted to search for the optimal transmitting and receiving beams with the maximum beamforming gain \cite{Ref:BEAM_TRAINING1,Ref:BEAM_TRAINING2,Ref:BEAM_TRAINING3,Ref:BEAM_TRAINING4,Ref:BEAM_TRAINING5}. Since the beams are generally selected from the predefined finite-size codebook, the brute-force beam search, which exhaustively sweeps all the transmitting and receiving beam pairs in the codebook, is an optimal beam training scheme \cite{Ref:BEAM_TRAINING1}. However, this may lead to excessively high training overhead. To address this problem, the two-level beam search scheme based on a hierarchical multi-resolution codebook was proposed in \cite{Ref:BEAM_TRAINING2,Ref:BEAM_TRAINING3}, where the first-level search aims to find the optimal wide beam, and the second-level search confirms the optimal narrow beam direction in the range of the selected wide beam. Another low-training-overhead scheme is the interactive beam search, where the candidate transmitting and receiving beams are swept separately \cite{Ref:BEAM_TRAINING4,Ref:BEAM_TRAINING5}. Nevertheless, the two-level and interactive beam search schemes still bring considerable training overhead.

Various schemes based on conventional beam search have been proposed to enhance beam alignment accuracy \cite{Ref:R3_2} or to reduce beam training overhead \cite{Ref:CALIBRATED_TRAINING, Ref:R4_1, Ref:R4_2}. An optimized two-stage search algorithm was proposed in \cite{Ref:R3_2} to better utilize the fixed training budget and ensure good alignment accuracy. The work \cite{Ref:CALIBRATED_TRAINING} proposed to calculate the optimal narrow beam based on the ratio of the beamforming gains between the selected wide beam and its neighboring wide beams, but this scheme is sensitive to noise and multipath interference. The study \cite{Ref:R4_1} proposed an interactive beam alignment procedure that minimizes BS transmit power subject to throughput constraints. Moreover, an adaptive sequential alignment algorithm based on the hierarchical codebooks was proposed in \cite{Ref:R4_2}, which selects the most likely beam based on the posterior distribution. However, this algorithm is prone to errors due to under-exploration of the beam space.

Recently, deep learning has been broadly adopted to enhance the performance of wireless communications \cite{Ref:DEEP_LEARNING1,Ref:DEEP_LEARNING2}. In order to reduce the overhead of beam training, deep learning was introduced to predict the optimal beam directly \cite{Ref:BEAM_PREDICTION1,Ref:BEAM_PREDICTION2,Ref:BEAM_PREDICTION3,Ref:BEAM_PREDICTION4,Ref:BEAM_PREDICTION5,Ref:BEAM_PREDICTION6,Ref:BEAM_PREDICTION7}. Specifically, a coordinated beamforming scheme was proposed in \cite{Ref:BEAM_PREDICTION1}, which uses convolutional neural network (CNN) to predict the optimal beam from the pilot signals received at multiple BSs. Deep neural network (DNN) was applied to determine mmWave beams based on the low-frequency channel state information (CSI) in \cite{Ref:BEAM_PREDICTION2,Ref:BEAM_PREDICTION3}. The works \cite{Ref:BEAM_PREDICTION4,Ref:BEAM_PREDICTION5} proposed to exploit cameras to assist the mmWave beam prediction based on deep learning tools. However, the studies \cite{Ref:BEAM_PREDICTION1,Ref:BEAM_PREDICTION2,Ref:BEAM_PREDICTION3,Ref:BEAM_PREDICTION4,Ref:BEAM_PREDICTION5} rely on auxiliary information, which may not be available in most scenarios. By contrast, the works \cite{Ref:BEAM_PREDICTION6,Ref:BEAM_PREDICTION7} utilized DNN to predict the optimal beam according to the training results of the uniformly sampled beams. However, noise and multipath interference can degrade the prediction accuracy seriously when the directions of the sampled beams are not adjacent to the dominant path.

Because of the relative stability of UE movement within a short time, prior channel information can be used to track UE locations and assist beam training \cite{Ref:BEAM_TRACKING1,Ref:BEAM_TRACKING2,Ref:BEAM_TRACKING3,Ref:BEAM_TRACKING4,Ref:R3_3,Ref:R1_1,Ref:R1_2,Ref:R4_3,Ref:BEAM_TRACKING5}. The extended Kalman filter (EKF) has been widely applied to track the angle of the dominant path \cite{Ref:BEAM_TRACKING1,Ref:BEAM_TRACKING2,Ref:BEAM_TRACKING3}, but this method suffers from error propagation according to \cite{Ref:R3_3}. In \cite{Ref:R3_3}, auxiliary beams were adopted for beam tracking, where two beams with perturbations from the former estimated angle are measured to track the angle variation. However, the method cannot be applied to the codebook based scenario, since the directions of auxiliary beams may not be perfectly matched. Differently, the works \cite{Ref:R1_1,Ref:R1_2} formulated the time-varying AoA and AoD as discrete Markov processes with known transition probabilities, and proposed beam tracking strategies to maximize the successful tracking probability. The BS handover was further taken into consideration to fight against beam misalignment and blockage in mobile scenarios \cite{Ref:R4_3}. Alternatively, long-short term memory (LSTM) network was adopted to infer the optimal mmWave beam at the target BS based on the prior CSI of the source mmWave BS \cite{Ref:BEAM_TRACKING5}. However, the estimation of mmWave CSI may lead to huge pilot overhead due to large number of antennas.

Motivated by the feasibility of estimating the angle of the dominant path based on the channel power leakage in the received signals of mmWave beam training \cite{Ref:BEAM_PREDICTION7,Ref:LEAKAGE1}, this paper proposes to calibrate the narrow beam direction by utilizing the received signals of wide beam training. Also the prior received signals are used to track the movement of UE, which can further reduce beam misalignment caused by noise. Considering the complex nonlinear properties of the channel power leakage, we adopt deep learning to predict the optimal narrow beam directly.

More specifically, three deep learning assisted calibrated beam training schemes are proposed. The first scheme leverages CNN to implement the prediction based on the instantaneous received signals of wide beam training. Since the prediction results are expressed as the probability that each candidate narrow beam is the optimal one, the additional narrow beam training according to the predicted probabilities can be performed to further calibrate beam directions. However, this scheme relies only on one wide beam training, which is sensitive to noise. To address this issue, in the second scheme, in order to enhance the robustness to noise, LSTM is utilized to track the movement of UE and calibrate the narrow beam direction according to the received signals of prior beam training. To further reduce the overhead of wide beam training, an adaptive beam training strategy is proposed in the third scheme, where partial wide beams are selected to be trained based on the received signals of prior beam training. Two criteria of the wide beam selection, namely, optimal neighboring criterion (ONC) and maximum probability criterion (MPC), are designed, where ONC selects the neighboring wide beams of the predicted optimal beam, while MPC selects the wide beams with the top predicted probabilities. Moreover, since the optimal beam direction may switch in mobile scenarios, auxiliary LSTM is introduced to predict the optimal wide beam corresponding to the current instant in advance for calibrating the directions of the selected wide beams more precisely. Simulation results demonstrate that our proposed schemes achieve significantly higher beamforming gain with smaller beam training overhead compared with the conventional and existing deep-learning based counterparts.

The main contributions of this paper can be summarized as follows:
\begin{itemize}
\item{We propose a wide beam based training method to predict the optimal narrow beam, where CNN is applied to implement the prediction.}
\item{We propose to enhance the prediction accuracy by using the received signals of prior beam training, where LSTM is applied to extract the UE movement information for further calibrating the predicted beam direction.}
\item We design an adaptive beam training strategy, where two criteria, ONC and MPC, are proposed to select partial wide beams to be trained based on the received signals of prior beam training. Moreover, we propose an auxiliary LSTM to calibrate the directions of the selected wide beams more precisely.
\end{itemize}

The paper is organized as follows. Section~\ref{Sec:SYS} presents the channel model and beam training model. Our three calibrated beam training schemes are detailed in Sections~\ref{Sec:SCHEME1}, \ref{Sec:SCHEME2} and \ref{Sec:SCHEME3}, respectively. Section~\ref{Sec:SIM} presents the simulation results. Our conclusions are drawn in Section~\ref{Sec:CONCLUSION}.

We adopt the following notational conventions. $\mathbb{Z}$ denotes the set of integers, $\mathbb{N}^*$ is the set of positive integers, and $\mathbb{C}^{m\times n}$ denotes the $m\times n$ complex space. Boldface capital and lower-case letters stand for matrices and vectors, respectively, e.g., $\bm{A}$ and $\bm{a}$, while calligraphic capital letters denote sets, e.g., $\mathcal{A}$. The logical AND is denote by $\wedge$, and $\textsf{j}=\sqrt{-1}$, while $\Re(\cdot)$ and $\Im(\cdot)$ denote the real and imaginary parts of a complex number, respectively. The transpose and conjugate transpose operators are denoted by $(\cdot )^{\text{T}}$ and $(\cdot )^{\text{H}}$, respectively, while $|\cdot|$ denotes the magnitude operator. The $n\times n$ identity matrix is denoted as $\bm{I}_n$ and $\bm{0}_n$ is the $n$-dimensional vector whose elements are all zero, while $\|\cdot\|_2$ and $\|\cdot\|_\infty$ denote the 2-norm and infinite norm, respectively. $\langle \cdot \rangle$ denotes the order statistics, e.g., for $\mathcal{A} = \{a_1, a_2, \ldots , a_n\}$, $\langle \mathcal{A} \rangle = \{a_{\sigma_1}, a_{\sigma_2}, ..., a_{\sigma_n}\}$ with $a_{\sigma_1} \leq a_{\sigma_2} \leq ... \leq a_{\sigma_n}$. The notation $\hat{}$ on the top of a variable indicates the estimated value, e.g., $\hat{p}$, and mod stands for modulo operator.

\section{System Model}\label{Sec:SYS} 

\subsection{Channel Model}\label{S2.1}

Consider the downlink mmWave multiple-input multiple-output (MIMO) communication system serving single user, where BS and UE are equipped with $M_{\text{Tx}}$ and $M_{\text{Rx}}$ antennas, respectively. Further assume that a single radio frequency (RF) chain is employed at both BS and UE sides. Since the line-of-sight (LOS) path is typically significant, exploiting the low-attenuation LOS path can efficiently enhance the coverage of mmWave signals \cite{Ref:LOS3,Ref:LOS4}. For simplicity, we assume the two-dimensional (2D) channel model, where only azimuth angles are considered.

We consider the narrowband frequency-flat channel model \cite{Ref:BEAM_TRACKING5,Ref:NARROWBAND1} consisting of the LOS path and $C$ clusters. Specifically, the channel matrix $\bm{H} \in \mathbb{C}^{M_{\text{Rx}} \times M_{\text{Tx}}}$ can be expressed as
\begin{align}\label{Eq:channel_model} 
\bm{H} =& \underbrace{\sqrt{\frac{M_{\text{Tx}}M_{\text{Rx}}}{\rho_{\text{LOS}}}} \alpha_{\text{LOS}} \bm{a}_{\text{Rx}} \big(\theta_{\text{LOS}}\big) \bm{a}_{\text{Tx}}^{\text{H}} \big(\phi_{\text{LOS}}\big)}_{\bm{H}_{\text{LOS}}} + \nonumber \\
 & \underbrace{\sum_{c=1}^{C}\sqrt{\frac{M_{\text{Tx}}M_{\text{Rx}}}{\rho_{c}}} \sum_{l=1}^{L_c} \frac{\alpha_{c,l}}{\sqrt{L_c}} \bm{a}_{\text{Rx}} \big(\theta_c + \theta_{c,l}\big) \bm{a}_{\text{Tx}}^{\text{H}} \big(\phi_c + \phi_{c,l}\big)}_{\bm{H}_{\text{NLOS}}} .
\end{align}
In this model, the $c$-th cluster containing $L_c$ paths has pathloss $\rho_c$, angle-of-arrival (AoA) $\theta_c$ and angle-of-departure (AoD) $\phi_c$, while $\alpha_{c,l}$, $\theta_{c,l}$ and $\phi_{c,l}$ are the complex gain, AoA offset and AoD offset, respectively, corresponding to the $l$-th path in the $c$-th cluster. Similarly, the LOS path has pathloss $\rho_{\text{LOS}}$, AoA $\theta_{\text{LOS}}$ and AoD $\phi_{\text{LOS}}$. For convenience, we use $\bm{H}_{\text{LOS}}$ and $\bm{H}_{\text{NLOS}}$ to represent the LOS part and the non-line-of-sight (NLOS) part of the channel matrix, respectively. Furthermore, $\bm{a}_{\text{Tx}} \in \mathbb{C}^{M_{\text{Tx}} \times 1}$ and $\bm{a}_{\text{Rx}} \in \mathbb{C}^{M_{\text{Rx}} \times 1}$ denote the antenna response vectors of BS and UE, respectively. We assume that uniform linear arrays (ULAs) are adopted at both BS and UE sides, and thus the two antenna response vectors are expressed respectively as
\begin{align} 
\bm{a}_{\text{Tx}} (\phi ) =& \sqrt{\frac{1}{M_{\text{Tx}}}} \big[1 ~ e^{\textsf{j}2\pi d_{\text{Tx}} \sin\phi/\lambda} \cdots e^{\textsf{j}2\pi (M_{\text{Tx}}-1) d_{\text{Tx}} \sin\phi / \lambda}\big]^{\text{T}}, \label{Eq:ULA_TX} \\
\bm{a}_{\text{Rx}} (\theta ) =& \sqrt{\frac{1}{M_{\text{Rx}}}} \big[1 ~ e^{\textsf{j}2\pi d_{\text{Rx}} \sin\theta/\lambda}\cdots e^{\textsf{j}2\pi (M_{\text{Rx}}-1) d_{\text{Rx}} \sin\theta/\lambda}\big]^{\text T}, \label{Eq:ULA_RX}
\end{align}
where $d_{\text{Tx}}$ and $d_{\text{Rx}}$ are the antenna spacings at BS and UE, respectively, $\lambda$ denotes the wavelength, $\phi$ and $\theta$ denote the corresponding AoD and AoA. For simplicity, we set $d_{\text{Tx}} = d_{\text{Rx}} = \lambda / 2$.

\subsection{Beam Training Model}\label{S2.2}

We assume that phase shifter based analog beamforming is applied, where $\bm{f} \in \mathbb{C}^{M_{\text{Tx}} \times 1}$ aligned with the direction $\gamma_{\text{Tx}}$ is denoted as the transmitting beam of BS, and $\bm{w} \in \mathbb{C}^{M_{\text{Rx}} \times 1}$ aligned with the direction $\gamma_{\text{Rx}}$ is denoted as the receiving beam of UE. The transmitting and receiving beams are selected from the predefined codebooks $\mathcal{F}$ and $\mathcal{W}$, which consist of $N_{\text{Tx}}$ and $N_{\text{Rx}}$ candidate beams, respectively. Assuming that the discrete Fourier transform (DFT) codebook is utilized \cite{Ref:DFT_CODEBOOK}, the candidate transmitting beam $\bm{f}_m$, $m \in \{1, 2, ..., N_{\text{Tx}}\}$, and receiving beam $\bm{w}_n$, $n \in \{1, 2, ..., N_{\text{Rx}}\}$, can be written respectively as
\begin{align} 
\bm{f}_m =& \sqrt{\frac{1}{M_{\text{Tx}}}}\big[1 ~ e^{\textsf{j} \pi \sin{\gamma_{\text{Tx},m}}}\cdots e^{\textsf{j} \pi (M_{\text{Tx}}-1) \sin{\gamma_{\text{Tx},m}}}\big]^{\text T}, \label{Eq:beam_TX} \\
\bm{w}_n =& \sqrt{\frac{1}{M_{\text{Rx}}}}\big[1 ~ e^{\textsf{j} \pi \sin{\gamma_{\text{Rx},n}}}\cdots e^{\textsf{j} \pi (M_{\text{Rx}}-1) \sin{\gamma_{\text{Rx},n}}}\big]^{\text T}, \label{Eq:beam_RX}
\end{align}
where $\gamma_{\text{Tx},m}$ and $\gamma_{\text{Rx},n}$ denote the beam directions of the $m$-th candidate beam at BS side and the $n$-th candidate beam at UE side, respectively. To cover the whole angular spaces of BS $\Gamma_{\text{Tx}}$ and UE $\Gamma_{\text{Rx}}$, we assume that the transmitting and receiving beam directions are uniformly sampled respectively in $\big(-\Gamma_{\text{Tx}}/2,~\Gamma_{\text{Tx}}/2\big)$ and $\big(-\Gamma_{\text{Rx}}/2,~ \Gamma_{\text{Rx}}/2\big)$ \cite{Ref:CALIBRATED_TRAINING}, i.e.,
\begin{align}
\gamma_{\text{Tx},m} =& -\frac{\Gamma_{\text{Tx}}}{2} + \frac{2m-1}{2N_{\text{Tx}}} \Gamma_{\text{Tx}}, \label{eq6} \\
\gamma_{\text{Rx},n} =& -\frac{\Gamma_{\text{Rx}}}{2} + \frac{2n-1}{2N_{\text{Rx}}} \Gamma_{\text{Rx}}. \label{eq7}
\end{align}
Given the channel matrix $\bm{H}$ and beam pair $\{\bm{f},\bm{w}\}$, the received signal $y$ can be written as
\begin{equation}\label{Eq:signal_model} 
y = \sqrt{P} \bm{w}^{\text{H}} \bm{H} \bm{f} x + \bm{w}^{\text{H}} \bm{n} ,
\end{equation}
where $P$ is the transmit power and $x$ is the transmitted signal with $|x|=1$, while $\bm{n} \in \mathbb{C}^{M_{\text{Rx}} \times 1}$ denotes the additional white Gaussian noise (AWGN) vector with the noise power $\sigma^2$, i.e., $\bm{n} \sim \mathcal{CN} \big(\bm{0}_{M_{\text{Rx}}},\sigma^2 \bm{I}_{M_{\text{Rx}}}\big)$.

Beam training aims to find the optimal beam pair $\{\bm{f}_{m^{\star}} , \bm{w}_{n^{\star}}\}$ with the maximum beamforming gain, which can be formulated as the following optimization problem
\begin{equation}\label{beam_pair_training} 
\{m^{\star}, n^{\star}\} = \arg \max\limits_{\substack{m \in \{1, 2, \ldots, N_{\text{Tx}}\},\\n \in \{1, 2, \ldots, N_{\text{Rx}}\}}}\big| \bm{w}^{\text{H}}_n \bm{H} \bm{f}_m\big|^2.
\end{equation}
A straightforward scheme of beam training to solve the above optimization is the brute-force beam search, where all the candidate transmitting and receiving beams are swept to find the beam pair with the maximum power of the received signal \cite{Ref:BEAM_TRAINING1}. However, the scheme requires $N_{\text{Tx}} N_{\text{Rx}}$ measurements, which leads to excessively huge training overhead.

To tackle this problem, the two-level beam search based on a hierarchical multi-resolution codebook can be considered, where the codebook consists of the wide beam codewords in the first level and the narrow beam codewords in the second level \cite{Ref:BEAM_TRAINING2, Ref:BEAM_TRAINING3}. To illustrate this approach, consider obtaining the wide beams by switching on partial antennas \cite{Ref:HIERARCHICAL_CODEBOOK}. Specifically, $M_{\text{Tx}}/s_{\text{Tx}}$ antennas are utilized to generate $N_{\text{Tx}}/s_{\text{Tx}}$ wide beams at BS, where $s_{\text{Tx}}\! \in\! \mathbb{N}^*$ defines the number of narrow beams within each wide beam. Similarly, the wide beams at UE can be implemented with $M_{\text{Rx}}/s_{\text{Rx}}$ antennas, where $s_{\text{Rx}}\! \in\! \mathbb{N}^*$ denotes the number of narrow beams within each wide beam for UE. Therefore, the candidate transmitting wide beam $\bm{f}_{\text{w},m}$, $m\! \in\! \{1, 2, \ldots, N_{\text{Tx}} / s_{\text{Tx}}\}$, and receiving wide beam $\bm{w}_{\text{w},n}$, $n\! \in\! \{ 1, 2,\ldots, N_{\text{Rx}} / s_{\text{Rx}}\}$, can be written as
\begin{align} 
\bm{f}_{\text{w},m} =& \sqrt{\frac{1}{M_{\text{Tx}} / s_{\text{Tx}}}}\big[1 ~ e^{\textsf{j}\pi \sin{\gamma^{\text{w}}_{\text{Tx},m}}}\cdots e^{\textsf{j}\pi (\frac{M_{\text{Tx}}}{s_{\text{Tx}}}-1) \sin{\gamma^{\text{w}}_{\text{Tx},m}}}\big]^{\text T}, \label{Eq:widebeam_TX} \\
\bm{w}_{\text{w},n} =& \sqrt{\frac{1}{M_{\text{Rx}} / s_{\text{Rx}}}}\big[1 ~ e^{\textsf{j}\pi \sin{\gamma^{\text{w}}_{\text{Rx},n}}}\cdots e^{\textsf{j}\pi (\frac{M_{\text{Rx}}}{s_{\text{Rx}}}-1) \sin{\gamma^{\text{w}}_{\text{Rx},n}}}\big]^{\text T}, \label{Eq:widebeam_RX}
\end{align}
where the beam directions of the $m$-th candidate wide beam at BS side $\gamma_{\text{Tx},m}^{\text{w}}$ and the $n$-th candidate wide beam at UE side $\gamma_{\text{Rx},n}^{\text{w}}$ can be expressed as
\begin{align}
\gamma_{\text{Tx},m}^\text{w} =& -\frac{\Gamma_{\text{Tx}}}{2} + \frac{2m-1}{2N_{\text{Tx}}} s_{\text{Tx}} \Gamma_{\text{Tx}}, \label{eq12} \\
\gamma_{\text{Rx},n}^\text{w} =& -\frac{\Gamma_{\text{Rx}}}{2} + \frac{2n-1}{2N_{\text{Rx}}} s_{\text{Rx}} \Gamma_{\text{Rx}}. \label{eq13}
\end{align}
Based on the hierarchical multi-resolution codebook, the beam search is divided into two levels. The first-level searches for coarse beam alignment based on the wide beam codebook, given by
\begin{equation}\label{eq14}
\big\{m^{\star}_\text{w}, n^{\star}_\text{w}\big\} =\arg \max\limits_{\substack{m\in \{1,2,\ldots, N_{\text{Tx}} / s_{\text{Tx}}\}, \\ n\in \{1,2,\ldots, N_{\text{Rx}} / s_{\text{Rx}}\}}}\big| \bm{w}_{\text{w},n}^{\text{H}} \bm{H}_\text{w} \bm{f}_{\text{w},m}\big|^2,
\end{equation}
where ${\bm{H}_\text{w}}\! \in\! \mathbb{C}^{M_{\text{Rx}} / s_{\text{Rx}} \times M_{\text{Tx}} / s_{\text{Tx}}}$ is the sub-channel matrix corresponding to the antennas for wide beam training. The first level search requires $N_{\text{Tx}} N_{\text{Rx}} / s_{\text{Tx}} s_{\text{Rx}}$ measurements. Recall that $N_\text{Tx}$ and $N_\text{Rx}$ antennas correspond to $N_\text{Tx}$ and $N_\text{Rx}$ candidate transmitting and receiving narrow beams of (\ref{Eq:beam_TX}) and (\ref{Eq:beam_RX}) with the directions of (\ref{eq6}) and (\ref{eq7}), respectively. The second-level search confirms the `optimal' narrow beam pair in the range of the selected wide beam pair (\ref{eq14}), given by
\begin{equation}\label{eq15}
\{m^{\star}, n^{\star}\} = \arg \max\limits_{\substack{m \in \{(m^{\star}_\text{w} - 1)s_{\text{Tx}} + 1,\ldots,  m^{\star}_\text{w}s_{\text{Tx}}\},\\ n \in \{(n^{\star}_\text{w} - 1)s_{\text{Rx}} + 1,\ldots, n^{\star}_\text{w} s_{\text{Rx}} \} }}\big| \bm{w}^{\text{H}}_n \bm{H} \bm{f}_m\big|^2.
\end{equation}
The second-level search needs further $s_{\text{Tx}} s_{\text{Rx}}$ measurements. Hence, the two-level beam search requires $N_{\text{Tx}} N_{\text{Rx}} / s_{\text{Tx}} s_{\text{Rx}} + s_{\text{Tx}} s_{\text{Rx}}$ measurements, which is significantly smaller than that imposed by the brute-force beam search.

Another overhead-reducing scheme is the interactive beam search, which selects the beams at BS and UE sides separately \cite{Ref:BEAM_TRAINING4, Ref:BEAM_TRAINING5}. Specifically, with UE antennas set to be the omni-directional pattern, BS sweeps all candidate transmitting beams to find the one with the maximum beamforming gain. Then with this `optimal' transmitting beam, UE sweeps all candidate receiving beams to find the beam with the maximum beamforming gain. In other words, the `optimal' beam pair are obtained by solving the following two optimization problems separately:
\begin{align} 
m^\star =& \arg \max\limits_{m\in \{1,2,\ldots,N_{\text{Tx}}\}} \big\|\bm{H} \bm{f}_m\big\|_{2}^2, \label{Eq:downlink_beam_training} \\
n^\star =& \arg \max\limits_{n\in \{1,2,\ldots,N_{\text{Rx}}\}} \big| \bm{w}^{\text{H}}_n \bm{H} \bm{f}_{m^\star}\big|^2. \label{Eq:uplink_beam_training}
\end{align}
This scheme requires $N_{\text{Tx}} + N_{\text{Rx}}$ measurements, which is much lower than the brute-force search.

\section{CNN Assisted Calibrated Beam Training}\label{Sec:SCHEME1} 

\subsection{Motivation}\label{S3.1}

As indicated in (\ref{Eq:downlink_beam_training}) and (\ref{Eq:uplink_beam_training}), the beam search can be implemented at BS and UE separately. For simplicity, we investigate the selection of the transmitting beams at BS side, where the single-antenna UE is assumed and thus the receiving beam $\bm{w}$ is omitted. Since mmWaves have weak penetration ability and significant reflecting power loss, the power of the LOS path is considerably higher than its NLOS counterparts. Hence the LOS path is dominant in mmWave channels \cite{Ref:LOS1,Ref:LOS2}. To achieve the maximum beamforming gain, the transmitting beam direction $\gamma_{\text{Tx}}$ should be aligned with the AoD of the LOS path $\phi_\text{LOS}$, while other NLOS paths can be treated as the noise. Specifically, we can rewrite the received signal model (\ref{Eq:signal_model}) as
\begin{align}\label{Eq:equivalent_noise} 
y =& \sqrt{P} \bm{H}_{\text{LOS}} \bm{f} x + \sqrt{P} \bm{H}_{\text{NLOS}} \bm{f} x + \bm{n} 
= \sqrt{P} \bm{H}_{\text{LOS}} \bm{f} x + \bm{n}_\text{eq} ,
\end{align}
where the equivalent noise $\bm{n}_\text{eq}=\sqrt{P} \bm{H}_{\text{NLOS}} \bm{f} x + \bm{n}$. Substituting Eq.\,(\ref{Eq:channel_model}) into (\ref{Eq:equivalent_noise}) yields
\begin{align}\label{eq19}
y =& \sqrt{\frac{M_{\text{Tx}}M_{\text{Rx}} P}{\rho_{\text{LOS}}}} \alpha_{\text{LOS}} \bm{a}_{\text{Tx}}^{\text{H}} ( \phi_{\text{LOS}} ) \bm{f} x + \bm{n}_\text{eq} \notag \\
=& \sqrt{\frac{M_{\text{Tx}}M_{\text{Rx}} P}{\rho_{\text{LOS}}}} \alpha_{\text{LOS}} q( \phi_{\text{LOS}} ) x + \bm{n}_\text{eq},
\end{align}
where $q(\phi_{\text{LOS}}) = \bm{a}_{\text{Tx}}^{\text{H}} ( \phi_{\text{LOS}} ) \bm{f}$ reflects the alignment degree between $\gamma_{\text{Tx}}$ and $\phi_{\text{LOS}}$, which determines the beamforming gain.

\begin{figure}[bp!]
\vspace*{-6mm}
\begin{center}
\includegraphics[width=.5\textwidth]{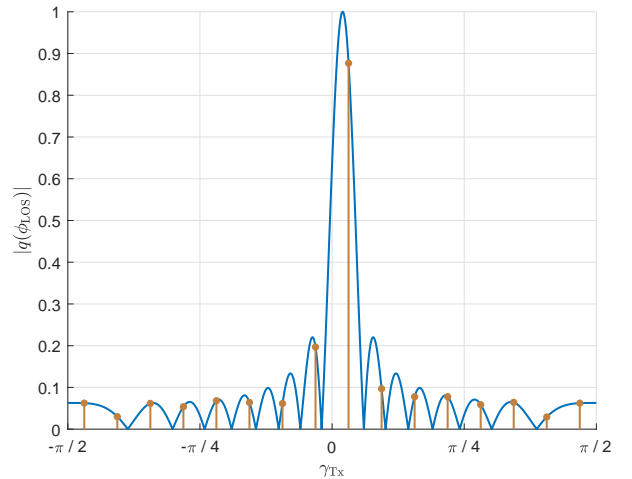}
\end{center}
\vspace*{-4mm}
\caption{Example of the channel power leakage, where the AoD of the LOS path $\phi_{\text{LOS}}  = 0.02 \pi$ and BS has $M_{\text{Tx}} = 16$ antennas. The blue curve assumes that the transmitting beam direction $\gamma_{\text{Tx}}$ is continuously distributed in $[-\pi / 2, ~ \pi / 2]$, while the brown dots show the results with $N_{\text{Tx}} = 16$ candidate transmitting beams.}
\label{Fig:leakage}
\vspace*{-1mm}
\end{figure}

Since the number of beam directions is limited under the on-grid assumption, the AoD of the LOS path $\phi_{\text{LOS}}$ may not be perfectly aligned, leading to the quantization error \cite{Ref:R1_2}. This error causes the channel power leakage in the received signals of beam training \cite{Ref:LEAKAGE1}. Specifically, assuming that the $m$-th candidate transmitting beam is applied, then $q_m(\phi_{\text{LOS}})=\bm{a}_{\text{Tx}}^{\text{H}} ( \phi_{\text{LOS}} ) \bm{f}_m$ is given by
\begin{align}\label{Eq:beamforming_gain} 
q_m(\phi_{\text{LOS}}) &= \frac{1}{M_{\text{Tx}}}\frac{\sin{\frac{\pi M_{\text{Tx}}\phi_m^\Delta }{2}}}{\sin{\frac{\pi \phi_m^\Delta }{2}}} e^{\textsf{j} \frac{\pi (M_{\text{Tx}} - 1) \phi_m^\Delta}{2}},
\end{align}
where $\phi_m^\Delta = \sin\gamma_{\text{Tx},m} - \sin\phi_{\text{LOS}}$.
If $\phi_{\text{LOS}}$ locates in the side lobe of the $m$-th candidate beam, i.e., $|\phi_m^\Delta| > 2/M_{\text{Tx}} \wedge \phi_m^\Delta \neq 2k/M_{\text{Tx}}$, $k \in \mathbb{Z}$, we have $|q_m(\phi_{\text{LOS}})| > 0$, which indicates that the power of the LOS path leaks to the $m$-th candidate beam. An example of the channel power leakage is illustrated in Fig.~\ref{Fig:leakage}, where it can be seen that the relative relations of the elements in $\bm{q}(\phi_{\text{LOS}}) = \big[q_1(\phi_{\text{LOS}}) ~ q_2(\phi_{\text{LOS}})\cdots q_{N_{\text{TX}}}(\phi_{\text{LOS}})\big]^{\text{T}}$ are decided by $\phi_{\text{LOS}}$. Further assume that the transmitted signal $x$ is fixed and the equivalent noise $\bm{n}_\text{eq}$ is omitted. Then the relative relations among the elements of $\bm{q}(\phi_{\text{LOS}})$ are reflected in the received signals, which provides the feasibility of estimating $\phi_{\text{LOS}}$ based on the received signals of beam training.

\subsection{Problem Formulation}\label{S3.2}

To reduce the beam training overhead, we propose to train a small number of candidate beams and calibrate the beam direction according to the received signals. How to find properly trained beams that achieve high accuracy under the given training overhead is crucial. Intuitively, a straightforward way is to uniformly sample partial beams \cite{Ref:BEAM_PREDICTION6,Ref:BEAM_PREDICTION7}, but its performance may degrade significantly due to the low signal-to-noise ratio (SNR) when the AoD of the LOS path $\phi_\text{LOS}$ does not locate in the main lobe of any sampled beam. Motivated by the two-level beam search with low training overhead where the wide beam codebook can cover the whole angular space, we propose to measure the received signals of wide beams for calibrating the beam direction. Different from \cite{Ref:R4_1} and \cite{Ref:R4_2}, the property of the channel power leakage is utilized to estimate the accurate AoD in our proposed approach. Specifically, the calibrated beam training scheme based on the wide beam codebook is proposed. For convenience, we define the received signal of the $m$-th candidate wide beam as $y_{{\text{w}},m}$ and concatenate the received signals of all the wide beams into the received signal vector $\bm{y}_{\text{w}}\! =\! \big[y_{{\text{w}},1} ~ y_{{\text{w}},2}\cdots y_{{\text{w},N_{\text{Tx}}}/{s_{\text{Tx}}}}\big]^{\text{T}}$. Since the narrow beam codebook enjoys higher angular resolution, the proposed calibrated beam training scheme aims to predict the index of the optimal narrow beam at BS side $m^\star$ based on the received signal vector of wide beams $\bm{y}_{\text{w}}$. Because the number of candidate narrow beams is limited, the prediction can be formulated as a multi-classification task, where each classified category corresponds to one candidate narrow beam. Mathematically, the prediction model can be represented by the classification function $f_1(\cdot)$ as
\begin{equation}\label{eq21}
m^\star = f_1(\bm{y}_{\text{w}}), m^\star \in \{ 1,2,...,N_\text{Tx} \}.
\end{equation}

However, it is difficult to implement this prediction by conventional estimation methods for two reasons. First, the relationship between $\bm{y}_\text{w}$ and $\phi_{\text{LOS}}$ is highly nonlinear, and second, the distribution of the equivalent noise $\bm{n}_\text{eq}$ is difficult to acquire since the NLOS paths vary with the propagation environment. These two reasons make the estimation too complicated by a conventional means. Consequently, deep learning with strong ability to learn complex nonlinear relations is utilized to implement the prediction \cite{Ref:FITTING}. Besides, we propose to perform the prediction at BS side, which has sufficient computational capability to ensure low prediction delay. Our proposed scheme consists of two stages, training and predicting. In the training stage, training data are collected to train the deep learning model, where each sample comprises a received signal vector as the model input and the index of the corresponding optimal narrow beam as the classification label, which can be obtained by conventional beam training schemes and fed back to BS. Because of the similar directional and power properties between uplink and downlink channels, the feedback overhead can be reduced by performing uplink beam training, where the received signal vector at BS side is used as the prediction input \cite{Ref:SIM1,Ref:SIM2}. After the model is well-trained with sufficient data, it switches to the prediction stage. In this stage, BS and UE only perform the wide beam search, and the corresponding received signals are leveraged to predict the optimal narrow beam by the well-trained model. Thus, the narrow beam search is avoided and the overhead of beam training is reduced considerably.

It is worth emphasizing that our scheme can be extended to various application scenarios. First, the scheme can be utilized to predict the optimal receiving narrow beam at the UE side with multiple antennas, since $\bm{w}^{\text{H}} \bm{a}_{\text{Rx}}(\theta_{\text{LOS}})$ can be analyzed in a similar manner to (\ref{Eq:beamforming_gain}). The scheme can be extended to the wideband multicarrier case, since the channel power leakage occurs in the received signals on subcarriers. The scheme can also be adopted in the multi-user scenario, where the calibration of beam directions is performed for each user separately. Our proposed scheme still works well in the NLOS scenario with one dominant cluster, e.g., the reconfigurable intelligent surface (RIS)-assisted scenario \cite{Ref:RIS1,Ref:RIS2}, where the beam direction is aligned with the AoD of this cluster while other clusters are treated as the noise.

\subsection{Model Design}\label{S3.3}

\begin{figure}[tp!]
\vspace*{-1mm}
\begin{center}
\includegraphics[width=.4\textwidth]{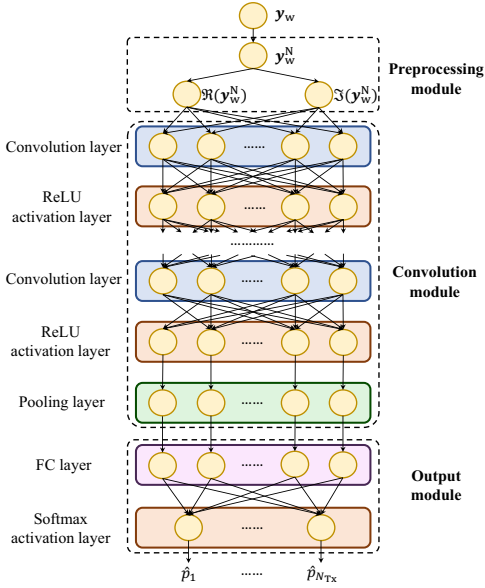}
\end{center}
\vspace{-4mm}
\caption{Proposed CNN based model, where each circle denotes one feature channel.}
\label{Fig:model1} 
\vspace{-4mm}
\end{figure}

CNN is adopted to implement the prediction due to its outstanding performance in classification tasks \cite{Ref:CNN}. The proposed CNN based model is depicted in Fig.~\ref{Fig:model1}, which can be divided into three parts, the preprocessing module, the convolution module and the output module.

\textbf{Preprocessing module}: Since the received signal vector $\bm{y}_\text{w}$ is complex-valued with large dynamic ranges, which cannot be fed to the CNN directly, the preprocessing module firstly normalizes $\bm{y}_\text{w}$ by the maximum amplitude of its elements, which can be written as
\begin{equation}\label{eq22}
\bm{y}_\text{w}^\text{N} = \frac{\bm{y}_\text{w} }{\| \bm{y}_\text{w}\|_\infty} .
\end{equation}
The normalized received signal vector $\bm{y}_\text{w}^\text{N}=\Re\big(\bm{y}_\text{w}^\text{N}\big)+\textsf{j}\Im\big(\bm{y}_\text{w}^\text{N}\big)$ is divided into the two real-valued feature channels of $\Re\big(\bm{y}_\text{w}^\text{N}\big)$ and $\Im\big(\bm{y}_\text{w}^\text{N}\big)$, which are fed to the following convolution module.

\textbf{Convolution module}: Multiple convolution layers are adopted to extract the hidden features from $\bm{y}_\text{w}^\text{N}$, where each layer is followed by the ReLU activation layer to provide nonlinear fitting ability. In order to avoid the overwhelmingly complex model, the pooling layer is introduced after the final ReLU activation layer, where each feature channel is downsampled to be a scalar.

\textbf{Output module}: To predict the optimal narrow beam from all the candidate narrow beams, the fully-connected (FC) layer is introduced after the pooling layer to implement the transformation from the extracted features to the candidate narrow beams, followed by a softmax activation layer for normalizing the outputs into probabilities, which can be written as
\begin{equation}\label{eq23}
\hat{p}_m =\text{softmax}\big(\bm{u}^{\text{T}}_m \bm{v} + b_m\big),m \in \{1,2,...,N_{\text{Tx}}\},
\end{equation}
where $\hat{p}_m$ is the predicted probability that the $m$-th candidate narrow beam is the optimal one, and $\bm{v}$ is the output vector of the pooling layer, while $\bm{u}_m$ and $b_m$ are the weight vector and bias of the FC layer corresponding to the $m$-th output. Finally, the narrow beam with the maximum predicted probability is selected, i.e.,
\begin{equation}\label{eq24}
\hat{m}^\star = \arg \max\limits_{m \in \{1, 2,\ldots , N_\text{Tx}\}} \hat{p}_{m} .
\end{equation}

The predicted probabilities provide the qualities of beams, and a beam with larger probability is predicted to enjoy higher beamforming gain over other beams with smaller probabilities. Therefore, the additional narrow beam training according to the predicted probabilities can be performed to further calibrate the beam directions. Specifically, the $K_\text{n}$ narrow beams with the top predicted probabilities are trained, whose indices $\mathcal{L}_{\text{n}}$ are specified by
\begin{align} 
&\big\{\hat{p}_{\sigma_1},\hat{p}_{\sigma_2},\cdots ,\hat{p}_{\sigma_{N_{\text{Tx}}}}\big\} = \langle \big\{ \hat{p}_1,\hat{p}_2,\cdots,\hat{p}_{N_{\text{Tx}}}\big\} \rangle , \label{eq25} \\
&\mathcal{L}_{\text{n}} = \big\{\sigma_{N_{\text{Tx}}-K_\text{n}+1}, \sigma_{N_{\text{TX}}-K_\text{n}+1},\cdots , \sigma_{N_{\text{Tx}}}\big\} . \label{eq26}
\end{align}
Let $y_m$ denote the received signal corresponding to the $m$-th candidate narrow beam. The narrow beam with the maximum power of the received signal is chosen as the optimal one, i.e.,
\begin{equation}\label{eq27}
\hat{m}^\star = \arg \max\limits_{m \in \mathcal{L}_{\text{n}}} |y_{m}|^2 .
\end{equation}
Obviously, increasing $K_\text{n}$ can enhance the beamforming gain at the cost of imposing higher training overhead.

Cross entropy loss is an evaluation metric widely used in classification tasks, which is utilized to train our proposed model. Mathematically, it can be expressed as
\begin{equation}\label{eq28}
\text{loss} = -\sum\limits_{m=1}^{N_{\text{Tx}}} p_m  \log\hat{p}_m ,
\end{equation}
where $p_m = 1$ if the $m$-th candidate narrow beam is the actual optimal beam. Otherwise $p_m =0$.

\section{LSTM Assisted Calibrated Beam Training}\label{Sec:SCHEME2}

\subsection{Motivation}\label{S4.1}

Although the scheme proposed in Section~\ref{Sec:SCHEME1} is capable of reducing the overhead of beam training, the prediction depends on the received signals of only one wide beam training, which lacks robustness to noise.
To address this problem, by exploiting the stability of UE movement within a short time, we can utilize prior information to track the UE movement and calculate the AoD of the LOS path $\phi_{\text{LOS}}$ based on the estimated UE location, such that the beam misalignment caused by noise can be calibrated. Beam training is periodically performed in mmWave communication systems, where typical training periods are smaller than 160\,ms \cite{Ref:TRAINING_PERIOD}. Because the received signals of beam training are the results of the interaction between the transmitted signals and the propagation environment around BS and UE, these signals manifest an RF signature of the UE location \cite{Ref:BEAM_PREDICTION1, Ref:FINGER_PRINT}. Consequently, prior received signals of beam training can be leveraged to track the UE movement and calibrate the beam direction without additional beam training overhead.

\subsection{Problem Formulation}\label{S4.2}

We now formulate the calibrated beam training scheme based on the received signals of prior beam training. Assume that beam training is performed periodically, and denote the received signals of the $t$-th wide beam training as $\bm{y}_{\text{w},t}$. To predict the optimal narrow beam corresponding to the $t$-th wide beam training $m^\star_t$, the received signals of both prior wide beam training $\big\{\bm{y}_{\text{w},1},\bm{y}_{\text{w},2}, \cdots , \bm{y}_{\text{w},t-1}\big\}$ and current wide beam training $\bm{y}_{\text{w},t}$ are jointly utilized. The prediction can be formulated as a multi-class classification task with the classification function $f_2(\cdot )$, i.e.,
\begin{equation}\label{eq29}
m^\star_t = f_2\big(\bm{y}_{\text{w}, 1},\bm{y}_{\text{w},2},\cdots , \bm{y}_{\text{w},t}\big), m^\star_{t} \in \{ 1,2,\cdots ,N_\text{Tx}\} .
\end{equation}
Since the AoD of the LOS path $\phi_{\text{LOS}}$ varies with the UE movement nonlinearly, we adopt deep learning model to implement the prediction.

Different from the model deployed in Section~\ref{Sec:SCHEME1}, in order to extract the UE movement features, the received signals of wide beam training and corresponding optimal narrow beam indices are packed in time order for UE, which forms a training sample.

\begin{figure*}[tp!]
\vspace*{-1mm}
\centering
	\subfigure[]{\label{Fig:LSTM1}
		\includegraphics[width=0.49\textwidth]{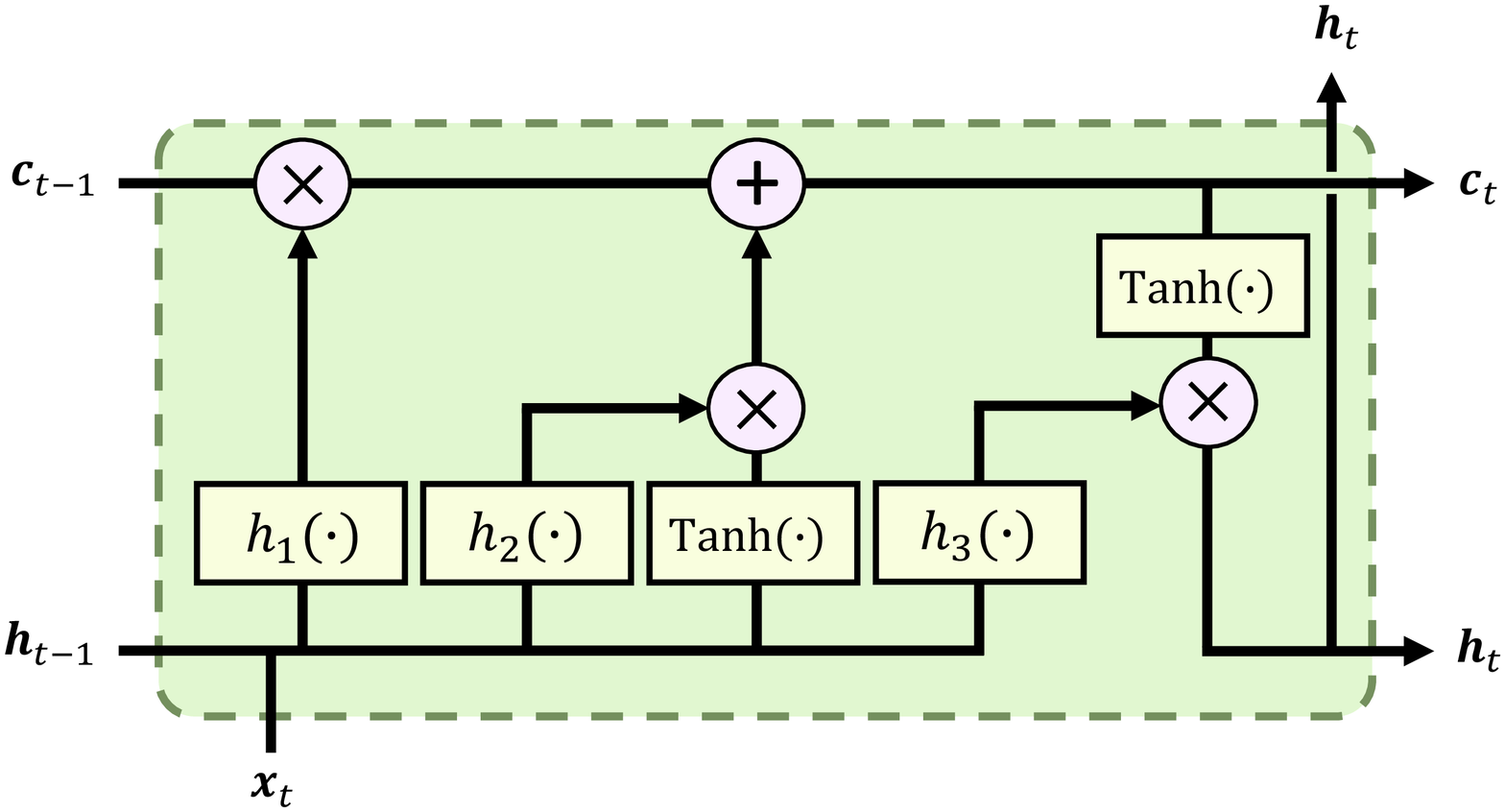}}
	\subfigure[]{\label{Fig:LSTM2}
		\includegraphics[width=0.49\textwidth]{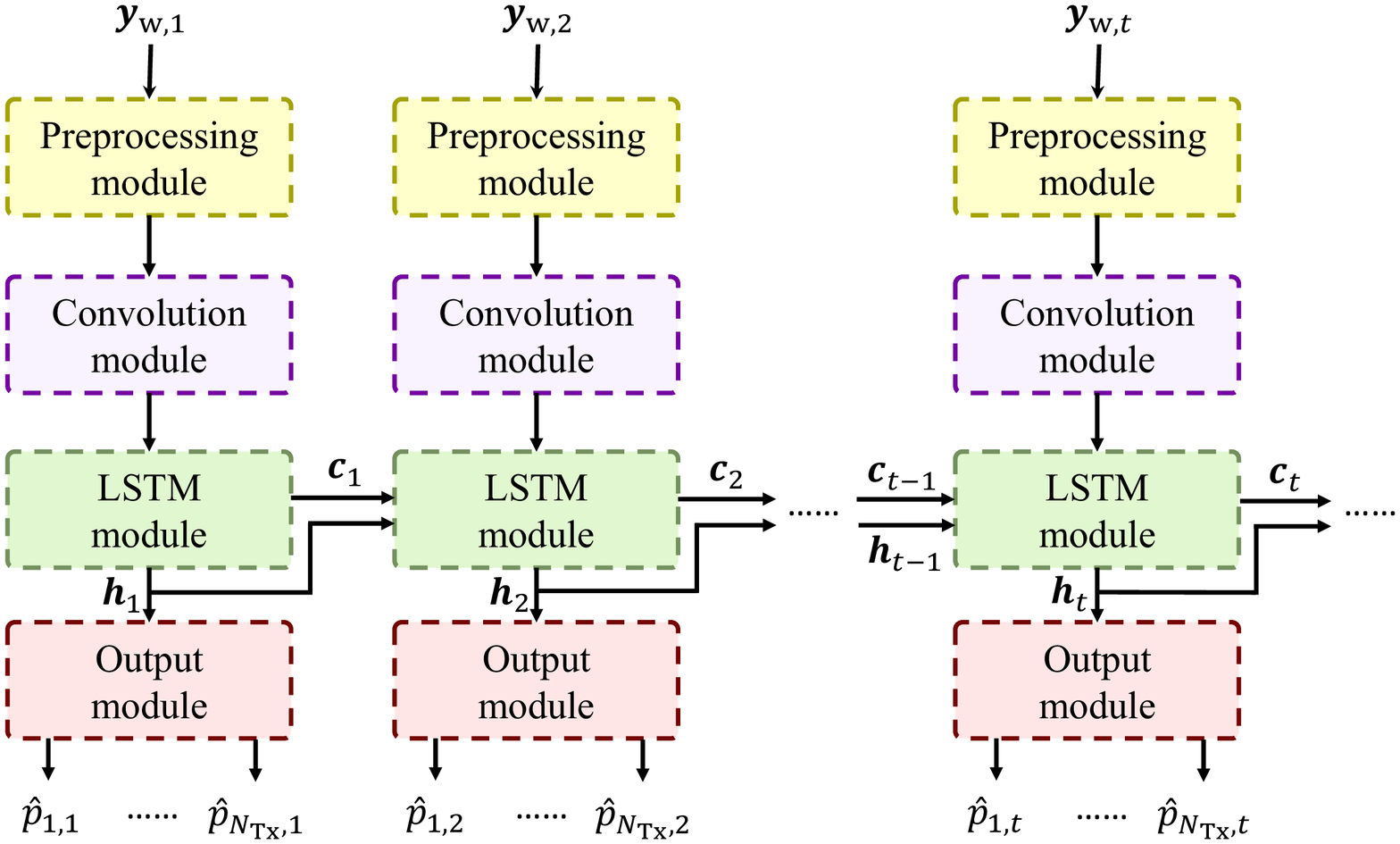}}
\vspace{-2mm}
\caption{(a)~Basic structure of LSTM, where $h_1(\cdot),h_2(\cdot)$ and $h_3(\cdot)$ denote hidden layers. (b) Proposed LSTM based model.} 
\vspace{-4mm}
\end{figure*}

\subsection{Model Design}\label{S4.3}

LSTM is used as the prediction model due to its excellent capability in temporal sequence learning \cite{Ref:LSTM}. The basic structure of LSTM is shown in Fig.~\ref{Fig:LSTM1}, where the input of the current time slot $\bm{x}_t$ together with the cell state and output of the previous time slot $\{\bm{c}_{t-1},\bm{h}_{t-1}\}$ are jointly fed to the LSTM at the $t$-th time slot, so that LSTM can learn the features from prior inputs. The proposed LSTM based model is depicted in Fig.~\ref{Fig:LSTM2}, which reuses partial structures in the CNN based model of Section~\ref{Sec:SCHEME1}. Once the $t$-th wide beam training is performed, the corresponding received signals are firstly fed to the preprocessing and convolution modules to extract the preliminary features related to $\bm{y}_{\text{w},t}$. Next, the LSTM module further calibrates the narrow beam direction based on the received signals of the current and prior beam trainings. Finally, the output module provides corresponding predicted probabilities $\big\{\hat{p}_{1,t},\hat{p}_{2,t},\cdots ,\hat{p}_{N_{\text{Tx}},t}\big\}$, and the narrow beam with the maximum probability is selected as the predicted optimal beam, whose index is denoted as $\hat{m}^\star_{t}$.

Cross entropy loss is also utilized to train the model, where the loss of one training sample is calculated as the average loss of all the narrow beam predictions for the UE.

\section{Adaptive Calibrated Beam Training}\label{Sec:SCHEME3} 

\subsection{Motivation}\label{S5.1}

The schemes proposed in Sections~\ref{Sec:SCHEME1} and \ref{Sec:SCHEME2} share one disadvantage that wide beam training still imposes considerable overhead. As illustrated in Fig.~\ref{Fig:leakage}, the leaked power of the beams far from the AoD of the LOS path $\phi_{\text{LOS}}$ is small, where effective information is difficult to extract from the corresponding received signals due to low SNRs. Therefore, we can train partial wide beams with high SNRs and use the corresponding received signals to predict the optimal narrow beam, such that the training overhead can be further reduced at the expense of slight degradation in beamforming gain. To find the high-SNR wide beams, the stability of UE movement again suggests that $\phi_{\text{LOS}}$ can be estimated from the received signals of prior beam training. Thus, we can select the wide beams to be trained based on the prior received signals adaptively.

\subsection{Basic Scheme}\label{Sec:SCHEME3_BASIC} 

Accordingly, we propose the adaptive calibrated beam training scheme, where partial wide beams are selected to be trained based on the received signals of prior beam training. To determine the initial AoD of the LOS path $\phi_{\text{LOS}}$ from the whole angular space, one full wide beam training is performed firstly, where the received signals of all the candidate wide beams are measured. Afterward, only partial wide beams need training, where the corresponding received signals are utilized to predict the optimal narrow beam index.

For the convenience of analysis, we focus on the beam selection for the $t$-th wide beam training with $t > 1$, and the corresponding AoD of the LOS path is denoted by $\phi_{\text{LOS}, t}$. Let $K$ be the number of the wide beams to be trained, and the corresponding indices are denoted as $\mathcal{L}_{\text{w},t}$. Two criteria are proposed to select the wide beams with high SNRs, which are the optimal neighboring criterion (ONC) and maximum probability criterion (MPC), respectively.

\textbf{ONC}: As indicated in Fig.~\ref{Fig:leakage}, the wide beam adjacent to the LOS path is more likely to enjoy high received power. Thus ONC aims to select the wide beams whose directions are the nearest to $\phi_{\text{LOS},t}$. Unfortunately, $\phi_{\text{LOS},t}$ cannot be accurately obtained. To find a proper approximation of $\phi_{\text{LOS}, t}$, it is noticed that the UE location at the $t$-th wide beam training is around the location corresponding to the $(t-1)$-th wide beam training, and consequently we propose to use the direction of the previous predicted optimal narrow beam $\gamma_{\text{Tx},\hat{m}^\star_{t-1}}$ to approximate $\phi_{\text{LOS},t}$. Mathematically, the ONC based beam selection is formulated as
\begin{align} 
&\gamma_{m,t}^\Delta = \big|\gamma_{\text{Tx},m}^\text{w} - \gamma_{\text{Tx},\hat{m}^\star_{t-1}}\big| \, \text{mod}\, \pi , \, m \in \left\{ 1, 2, \cdots,  \frac{N_{\text{Tx}}}{s_{\text{Tx}}} \right\}, \label{Eq:ONC_1} \\
& \left\{\gamma_{\sigma_1,t}^\Delta,\gamma_{\sigma_2,t}^\Delta,\cdots ,\gamma_{\sigma_{\frac{N_{\text{Tx}}}{s_{\text{Tx}}}},t}^\Delta\right\} = \left\langle \left\{ \gamma_{1,t}^\Delta, \gamma_{2,t}^\Delta,\cdots, \gamma_{\frac{N_{\text{Tx}}}{s_{\text{Tx}}},t}^\Delta\right\} \right\rangle, \label{Eq:ONC_2} \\
&\mathcal{L}_{\text{w},t} = \{ \sigma_1, \sigma_2,\cdots , \sigma_{K} \}. \label{Eq:ONC_3}
\end{align}
The direction difference $\big|\gamma_{\text{Tx},m}^\text{w} - \gamma_{\text{Tx},\hat{m}^\star_{t-1}}\big|$ modulo $\pi$ as expressed in (\ref{Eq:ONC_1}) is because $|q(\phi_{\text{LOS}})|$ is periodic with period $\pi$.

\textbf{MPC}: MPC is based on the property that the predicted probabilities reflect beam qualities, and it selects the wide beams with the top predicted probabilities in the $(t-1)$-th beam prediction. However, the prediction results only provide the probabilities of narrow beams instead of wide beams. To obtain the approximation of the predicted probability for the $m$-th wide beam $\hat{p}_{m,t}^{\text{w}_\text{ap}}$, the predicted probabilities of all the narrow beams within the $m$-th wide beam are added together. Therefore, the MPC based beam selection is mathematically formulated as
\begin{align} 
&\hat{p}_{m,t}^{\text{w}_\text{ap}} = \sum_{s=1}^{s_{\text{Tx}}} \hat{p}_{(m-1) s_{\text{Tx}}+s,t-1}, \, m\in \left\{1, 2,\cdots ,\frac{N_{\text{Tx}}}{s_{\text{Tx}}}\right\}, \label{Eq:MPC_1} \\
&\left\{ \hat{p}_{\sigma_1,t}^{\text{w}_\text{ap}},\hat{p}_{,\sigma_2,t}^{\text{w}_\text{ap}},\cdots ,\hat{p}_{\sigma_{\frac{N_{\text{Tx}}}{s_{\text{Tx}}}},t}^{\text{w}_\text{ap}}\right\} = \left\langle \left\{ \hat{p}_{1,t}^{\text{w}_\text{ap}},\hat{p}_{2,t}^{\text{w}_\text{ap}},\cdots ,\hat{p}_{\frac{N_{\text{Tx}}}{s_{\text{Tx}}},t}^{\text{w}_\text{ap}} \right\} \right\rangle, \label{Eq:MPC_2} \\
&\mathcal{L}_{\text{w},t} = \left\{ \sigma_{\frac{N_{\text{Tx}}}{s_{\text{Tx}}} - K + 1}, \sigma_{\frac{N_{\text{Tx}}}{s_{\text{Tx}}} - K + 2}, \cdots , \sigma_{\frac{N_{\text{Tx}}}{s_{\text{Tx}}}} \right\} . \label{Eq:MPC_3}
\end{align}

Once the $t$-th partial wide beam training is performed, the corresponding received signal vector $\bm{y}_{\text{wp},t}=\big[\bm{y}_{\text{wp},t}[1]\cdots \bm{y}_{\text{wp},t}[N_\text{Tx}/s_\text{Tx}]\big]^{\text{T}}\in \mathbb{C}^{(N_\text{Tx}/s_\text{Tx})\times 1}$ can be obtained according to
\begin{equation}\label{eq36}
\bm{y}_{\text{wp},t}[m]=\left\{ \begin{array}{cl}
 y_{m,t}^{\text{w}}, & \text{if } m \in \mathcal{L}_{\text{w},t}, \\
 0, & \text{otherwise},
\end{array} \right.
\end{equation}
$\forall m\in\{1,2,\cdots, N_\text{Tx}/s_\text{Tx}\}$, where $y_{m,t}^{\text{w}}$ is the received signal of the $m$-th candidate wide beam at the $t$-th wide beam training. Note that the format of $\bm{y}_{\text{wp},t}$ is similar to the received signal vector of full wide beam training $\bm{y}_{\text{w},t}$, and hence the received signal vectors of both the full and partial wide beam training can be processed by the same model. Therefore, the prediction model represented by the underlying multi-classification function $f_3(\cdot )$ is formulated as
\begin{equation}\label{eq37}
m^\star_{t} = f_3(\bm{y}_{\text{w}, 1}, \bm{y}_{\text{wp}, 2}, ..., \bm{y}_{\text{wp},t}), m^\star_{t} \in \{ 1,2,...,N_\text{Tx} \}.
\end{equation}

\begin{figure}[tp!]
\vspace*{-1mm}
\begin{center}
\includegraphics[width=.5\textwidth]{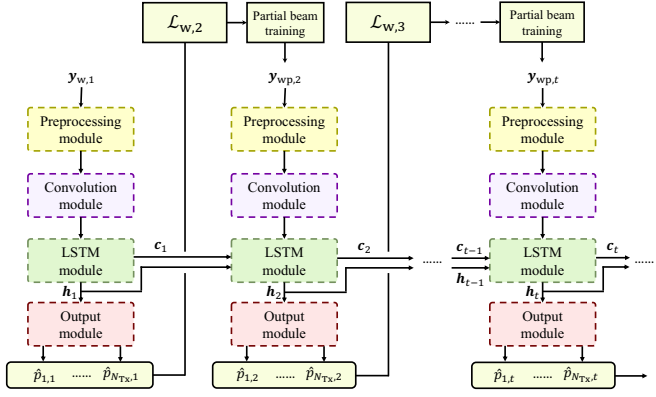}
\end{center}
\vspace{-4mm}
\caption{Proposed adaptive calibrated beam training model.}
\label{Fig:model3} 
\vspace*{-4mm}
\end{figure}

The proposed adaptive calibrated beam training model is illustrated in Fig.~\ref{Fig:model3}. The prediction model structures and loss function are the same as their counterparts in Section~\ref{Sec:SCHEME2}, but the model deployment is different from the previous case. Specifically, in the training stage, in order to learn to predict the optimal narrow beam from the partial received signals, the model simulates partial wide beam training, i.e., only the received signals of the selected wide beams are used as the model input. In the predicting stage, the proposed scheme only trains the selected wide beams and uses corresponding received signals to predict the optimal narrow beam, so that the overhead of wide beam training is significantly reduced.

\subsection{Enhanced Adaptive Calibrated Beam Training with Auxiliary LSTM}\label{Sec:SCHEME3_ENHANCED} 

\begin{figure*}[tp!]
\vspace*{-1mm}
\begin{center}
\includegraphics[width=.8\textwidth]{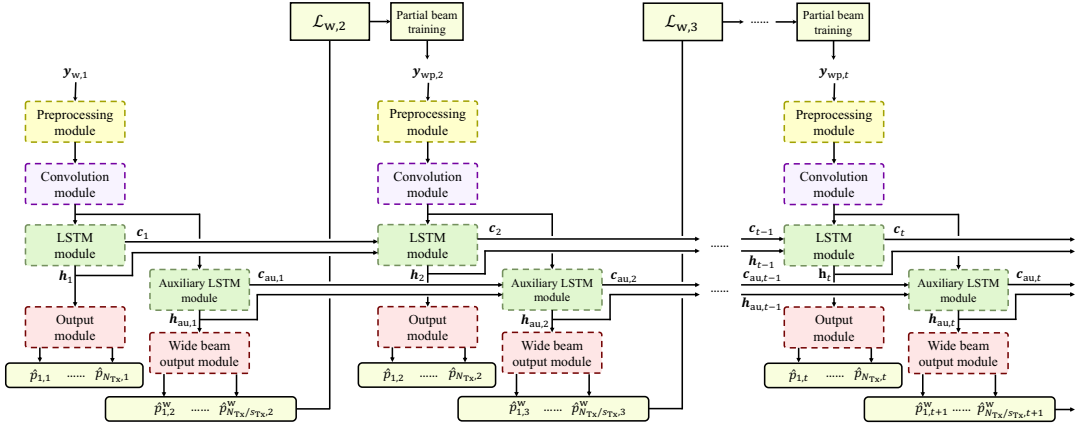}
\end{center}
\vspace{-4mm}
\caption{Proposed enhanced adaptive calibrated beam training model, where $\bm{c}_{\text{au},t}$ and $\bm{h}_{\text{au},t}$ are the cell state and output of the $t$-th auxiliary LSTM, respectively.}
\label{Fig:model4} 
\vspace{-4mm}
\end{figure*}

The proposed scheme of Subsection~\ref{Sec:SCHEME3_BASIC} selects the wide beams to be trained according to the results of the previous beam prediction. Therefore, the prediction may become outdated if the AoD of the LOS path $\phi_{\text{LOS}}$ varies in mobile scenarios. To track the varying AoD of the LOS path $\phi_{\text{LOS}}$, we can utilize the received signals of previous beam training to estimate the current UE location in advance, such that the indices of the selected wide beams to be trained can be calibrated to further enhance received SNRs.

Specifically, we propose the enhanced adaptive calibrated beam training scheme. To find the wide beams with high SNRs corresponding to the $t$-th wide beam training, auxiliary LSTM is introduced to predict the $t$-th optimal wide beam index $m^\star_{\text{w},t}$ in advance based on the received signals of prior wide beam training $\{{\bm{y}_{\text{w},1}, \bm{y}_{\text{wp},2},\cdots, \bm{y}_{\text{wp},t-1}}\}$, which can be expressed as
\begin{equation}\label{eq38}
m^\star_{\text{w},t} = f_{\text{au}}(\bm{y}_{\text{w},1}, \bm{y}_{\text{wp},2},\cdots, \bm{y}_{\text{wp},t-1}),\, m^\star_{\text{w},t} \in \left\{1,2,\cdots,\frac{N_\text{Tx}}{s_\text{Tx}}\right\},
\end{equation}
where $f_{\text{au}}(\cdot)$ denotes the above multi-classification function for the wide beam prediction. Similarly, the output of the model is expressed as the predicted wide beam probabilities $\{\hat{p}_{\text{1}, t}^{\text{w}}, \hat{p}_{\text{2}, t}^{\text{w}},\cdots,$ $\hat{p}_{N_{\text{Tx}}/s_\text{Tx},t}^{\text{w}}\}$, where the wide beam with the maximum predicted probability is selected as the optimal wide beam $\hat{m}^\star_{\text{w},t}$. In particular, ONC selects the wide beams whose directions are the nearest to the predicted optimal wide beam, and (\ref{Eq:ONC_1}) becomes
\begin{align}\label{eq39}
\gamma_{m,t}^\Delta = \big|\gamma_{\text{TX},m}^{\text{w}} - \gamma^{\text{w}}_{\text{TX},\hat{m}_{\text{w},t}^\star}\big|\, \text{mod}\, \pi, \,m \in \left\{1,2,\cdots, \frac{N_{\text{Tx}}}{s_{\text{Tx}}}\right\}.
\end{align}
On the other hand, MPC selects the wide beams with the top predicted probabilities, and (\ref{Eq:MPC_2}) is rewritten as
\begin{align}\label{eq40}
\left\{\hat{p}_{\sigma_1,t}^\text{w},\hat{p}_{\sigma_2,t}^\text{w},\cdots,\hat{p}_{\sigma_{\frac{N_{\text{Tx}}}{s_{\text{Tx}}}},t}^\text{w}\right\} = \left\langle \left\{ \hat{p}_{1,t}^\text{w},\hat{p}_{2,t}^\text{w},\cdots,\hat{p}_{\frac{N_{\text{Tx}}}{s_{\text{Tx}}},t}^\text{w} \right\} \right\rangle.
\end{align}

The enhanced adaptive calibrated beam training model is depicted in Fig.~\ref{Fig:model4}, where the LSTM module and the proposed auxiliary LSTM module share the same preprocessing and convolution modules to reduce the model overhead, and the wide beam output module is used to obtain the predicted probabilities of the wide beam prediction. Auxiliary LSTM does not require to collect extra training data in the training stage. Once the $t$-th wide beam training is performed, the wide beam index with the maximum power of the received signal is used as the classification label of the wide beam prediction.

Similarly, cross entropy loss is used as the loss function in wide beam predictions. To train the whole model with auxiliary LSTM, the losses of both narrow beam predictions $\text{loss}_{\text{n}}$ and wide beam predictions $\text{loss}_{\text{w}}$ are combined with the weight coefficient $\mu$, which can be expressed as
\begin{equation}\label{eq41}
\text{loss} = \text{loss}_{\text{n}} + \mu \text{loss}_{\text{w}} .
\end{equation}

\section{Simulation Study}\label{Sec:SIM} 

\subsection{Simulation System Setup}\label{S6.1}

{\renewcommand\arraystretch{0.2}
\begin{table*}[tp!]\scriptsize
\begin{minipage}{\textwidth}
\begin{minipage}[tp!]{0.4\textwidth}
\centering
\renewcommand\arraystretch{1.04}
\makeatletter\def\@captype{table}\makeatother\caption{Default system parameters.}\label{Tab:system_prameter}
\vspace{-2mm}
\begin{tabular}{cccc}
\toprule[0.8pt]
\rule{0pt}{7pt}
\footnotesize{Parameter} & \footnotesize{Value} \\
\toprule[0.8pt]
Center frequency $f_c$                                          & $28$\,GHz               \\
Bandwidth $W$                                                   & $2$\,MHz                \\
Group number of far scatterers $n_\text{g}$                     & $15$                    \\
Path number within one cluster $L_c$                            & $20$                    \\
Visible region radius $r_\text{v}$                              & $40$\,m                 \\
AoD spread within one cluster $\Delta \phi_\text{c}$            & $2.4^\circ$             \\
Delay spread within one cluster $\Delta \tau_\text{c}$          & $5$\,ns                 \\
Shadow fading standard deviation $\sigma_{\text{SF}}$           & $4$\,dB                 \\
Ricean K factor $K_{\text{R}}$                                  & $8$\,dB                 \\
BS antenna number $M_{\text{Tx}}$                               & $64$                    \\
BS wide beam number $N_{\text{Tx}} / s_{\text{Tx}}$             & $16$                    \\
BS narrow beam number $N_{\text{Tx}}$                           & $64$                    \\
Beam training period $\tau$                                     & $100$\,ms               \\
Total beam training number $T$                                  & $10$                    \\
Cell radius $r$                                                 & $100$\,m                \\
Transmit power $P$                                              & $15$\,dBm               \\
\toprule[0.8pt]
\end{tabular}
\end{minipage}
\begin{minipage}[tp!]{0.7\textwidth}
\centering
\renewcommand\arraystretch{1.13}
\makeatletter\def\@captype{table}\makeatother\caption{Structures and parameters of proposed deep learning models.}\label{Tab:model_prameter}
\vspace{-2mm}
\begin{tabular}{cccc}
\toprule[0.8pt]
\rule{0pt}{7pt}
\footnotesize{Module}                                        & \footnotesize{Layer} & \footnotesize{Parameter} \\
\toprule[0.8pt]
\multirow{5}{*}{\tabincell{c}{Convolution\\module}}          & Convolution          & $f_{i}=2,f_{o}=64,(3,3,1)$, BN                     \\

																														 & ReLU activation      & $f_{i}=64,f_{o}=64$ \\
                                                             & Convolution          & $f_{i}=64,f_{o}=256,(3,1,1)$, BN                    \\
                                                             & ReLU activation      & $f_{i}=256,f_{o}=256$ \\
                                                             & Pooling              & $f_{i}=256,f_{o}=256, \text{max-pooling}$           \\
\toprule[0.8pt]
\multirow{2}{*}{{\tabincell{c}{LSTM\\module}}}               & LSTM                 & $f_{i}=256,f_{o}=256,\text{dropout}=0.2$             \\
                                                             & LSTM                 & $f_{i}=256,f_{o}=256,\text{dropout}=0.2$             \\
\toprule[0.8pt]
\multirow{2}{*}{{\tabincell{c}{Auxiliary \\LSTM module}}}    & LSTM                 & $f_{i}=256,f_{o}=256,\text{dropout}=0.2$             \\
                                                             & LSTM                 & $f_{i}=256,f_{o}=256,\text{dropout}=0.2$             \\
\toprule[0.8pt]
\multirow{2}{*}{{\tabincell{c}{Output \\ module}}}           & FC                   & $f_{i}=256,f_{o}=64,\text{dropout}=0.3$             \\
                                                             & Softmax activation   & $f_{i}=64,f_{o}=64$ \\
\toprule[0.8pt]
\multirow{2}{*}{{\tabincell{c}{Wide beam \\ output module}}} & FC                   & $f_{i}=256,f_{o}=16,\text{dropout}=0.3$             \\
                                                             & Softmax activation   & $f_{i}=16,f_{o}=16$ \\
\toprule[0.8pt]
\end{tabular}
\end{minipage}
\end{minipage}
\vspace{-4mm}
\end{table*}
}

We consider a mmWave wireless communication system with LOS serving one single-antenna user, and the mobile scenario is assumed. Unless otherwise stated, UE performs the rectilinear motion with uniformly randomly distributed speed $v_{\text{UE}}\in[10, ~ 50]$\,m/s and acceleration $a_{\text{UE}}\in[-8, ~8]\,\text{m/s}^2$, and the motion direction is randomly generated in $[0, ~ 2\pi]$. In order to simulate the channel variations with UE movement, we apply the COST 2100 channel model \cite{Ref:COST2100, Ref:COST2100_CODE}, which defines several groups of far scatterers in the space and each group corresponds to one NLOS cluster together with a visible region, i.e., the area where the cluster exists. Based on the locations of BS, UE and far scatterers, the channel matrix $\bm{H}$ can be generated by (\ref{Eq:channel_model}). The default parameters of the simulated mmWave communication system are listed in Table~\ref{Tab:system_prameter}. The power of the AWGN $\sigma^2$ is calculated as $(-174 + 10\log_{10}W + N_{\text{F}})\,\text{dBm}$, where the noise factor $N_{\text{F}} = 6\,\text{dB}$. Moreover, the pathloss $\text{PL}$ is obtained as $\text{PL} = (26\log_{10}{d} + 20 \log_{10}{f_c}-147.56)\,\text{dB}$ \cite{Ref:COST2100_CODE}, where $d$ denotes the propagation distance.

For the proposed deep learning models, the detailed structures and parameters are specified in Table~\ref{Tab:model_prameter}, where $f_i$ and $f_o$ denote the numbers of input feature channels and output feature channels, respectively. The parameters in convolution layers $(p_1,p_2,p_3)$ represent the kernel size, sampling stride and zero-padding size, respectively. To accelerate model training convergence, batch normalization (BN) is applied in the convolution module, which transforms the processed data to the standard distribution with mean 0 and variance 1 \cite{Ref:BATCH_NORM}. To enhance model generalization ability, LSTM layers and FC layers exploit dropout strategy to abandon part of neurons randomly in the training stage for preventing overfitting \cite{Ref:DROPOUT}. We construct a dataset with 20,480 samples, where 80\% and 20\% of the dataset are used as the training set and the validation set, respectively. The model is trained for 80 epochs in the training stage, where Adam optimizer based on the back propagation algorithm is used to optimize the model parameters \cite{Ref:ADAM}.

\begin{figure*}[tp!] 
\vspace{-1mm}
\textcolor[rgb]{0.00,0.00,0.00}{
	\centering
	\subfigure[CNN assisted CBT of Sec. III.]{
		\includegraphics[width=0.499\textwidth]{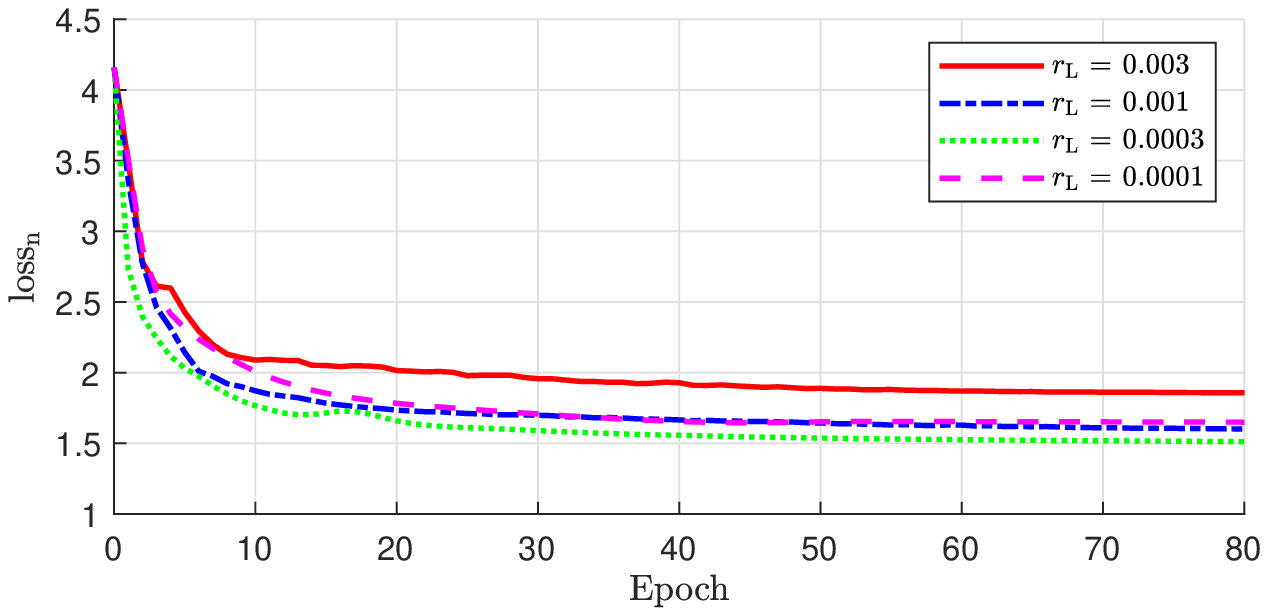}}
	\subfigure[LSTM assisted CBT of Sec. IV.]{
		\includegraphics[width=0.499\textwidth]{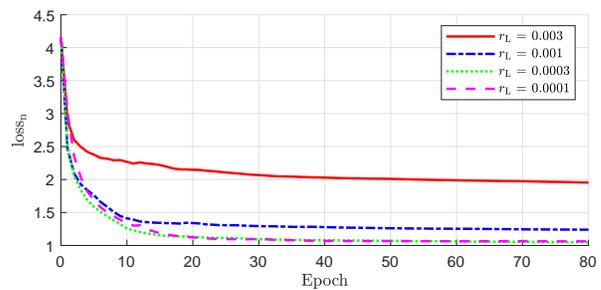}}}
    \vspace{-2mm}
	\textcolor[rgb]{0.00,0.00,0.00}{\subfigure[Adaptive CBT of Sec. V.]{
		\includegraphics[width=0.499\textwidth]{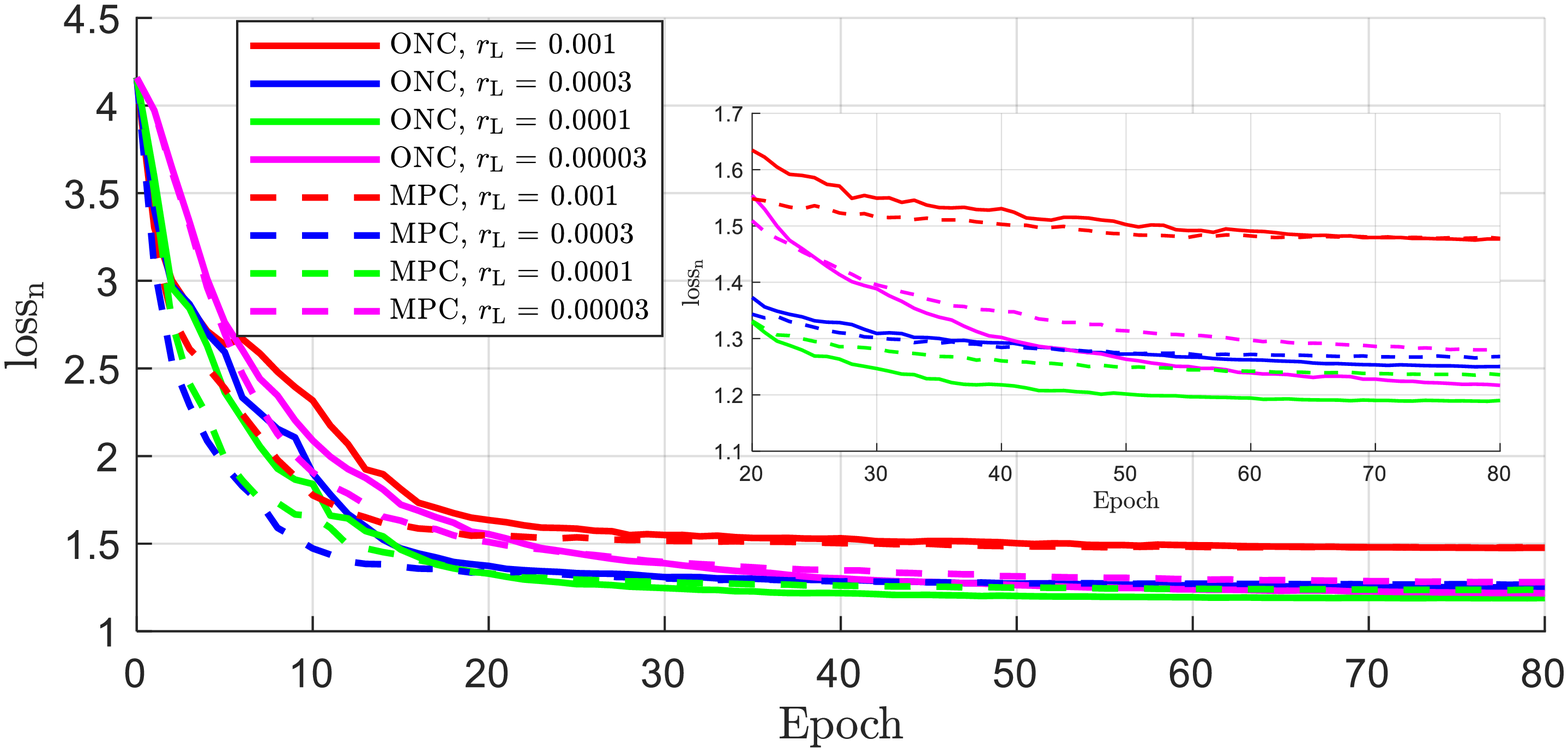}}
	\subfigure[Enhanced adaptive CBT of Sec. V with $K=7$ and $\mu = 1$.]{
		\includegraphics[width=0.499\textwidth]{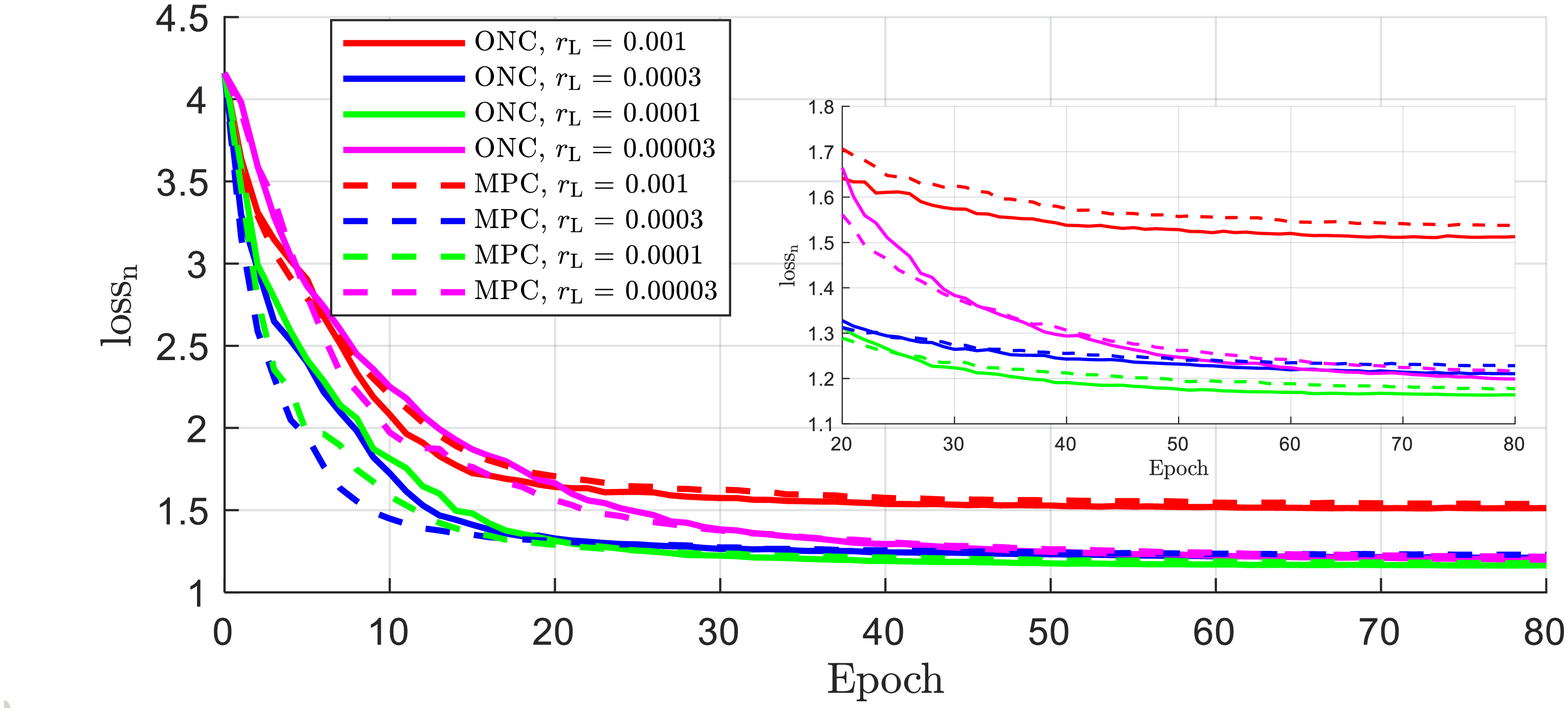}}
\caption{Narrow beam prediction loss as function of training epoch for proposed CBT schemes given different learning rates.}
\label{fig6}}
\vspace{-4mm}
\end{figure*}

Three metrics specified below are utilized for performance evaluation.
{\newline{1)~Cross entropy loss of the narrow beam prediction $\text{loss}_\text{n}$ defined in (\ref{eq28}), where the subscript denotes narrow beam prediction.}}
{\newline{2)~Normalized beamforming gain $G_{\text{N}}$ defined as
\begin{equation}\label{eq42}
G_{\text{N}} = \frac{|\bm{H} \bm{f}_{\hat{m}^\star}|^2}{|\bm{H} \bm{f}_{m^\star}|^2},
\end{equation}
where $\bm{f}_{m^\star}$ and $\bm{f}_{\hat{m}^\star}$ are the actual optimal narrow beam and the predicted optimal narrow beam, respectively.}}
{\newline{3)~Effective spectral efficiency $\overline{E}$ \cite{Ref:R4_2,Ref:R4_3} defined as
\begin{equation}\label{eq43}
\overline{E} = \frac{T_{\rm tot} - T_{\rm tra}}{T_{\rm tot}} \log_2 \left({1 + \frac{P|\bm{H}\bm{f}_{\hat{m}^\star}|^2}{\sigma^2}} \right) ,
\end{equation}
where $T_{\rm tot}$ is the total time of a communication session, and $T_{\rm tra}$ is the beam training time. Typically, beam training is performed periodically with the period $\tau$. Hence, $T_{\rm tra}$ equals to the product of $\frac{T_{\rm tot}}{\tau}$ with the number of trained beams required in each beam training and the duration of one beam measurement $t_{\rm s}$. In the simulation, we assume that $t_{\rm s}=0.1$\,ms \cite{Ref:R4_3}.

The average result over the entire validation set and 5 training runs is actually used as the evaluation metric value. The source code of our simulations can be found in \cite{Ref:SOURCE_CODE}.

\subsection{Investigation of Training Parameters and Complexity}\label{S6.2}

We now investigate the impact of the training parameters on the narrow beam prediction loss $\text{loss}_\text{n}$ for the proposed calibrated beam training (CBT) schemes.

The learning rate $r_\text{L}$ is a key algorithmic parameter of Adam optimizer \cite{Ref:ADAM}. Fig.~\ref{fig6} investigates the impact of $r_\text{L}$ on the achievable $\text{loss}_{\text{n}}$ for our proposed schemes. The number of trained wide beams in the enhanced adaptive CBT scheme is $K\! =\! 7$ with the weight coefficient $\mu \! =\! 1$. As expected, larger $r_\text{L}$ leads to higher $\text{loss}_{\text{n}}$ but faster convergence. It can be seen from Fig.~\ref{fig6} that to achieve a best combined steady state $\text{loss}_\text{n}$ performance and convergence rate, $r_\text{L}\! =\! 0.0003$ is appropriate for the CNN and LSTM assisted CBT schemes, while $r_\text{L}\! =\! 0.0001$ is appropriate for the adaptive CBT and enhanced adaptive CBT schemes.

Next, the impact of the weight coefficient $\mu$ on $\text{loss}_{\text{n}}$ for the enhanced adaptive CBT scheme under $K=7$ is investigated in Fig.~\ref{fig7}. It can be seen that $\text{loss}_{\text{n}}$ is minimized when $\mu$ is around $1$ for both the ONC and MPC based schemes. Therefore, $\mu$ is set to be $1$ in our simulation study.

\begin{figure}[tp!] 
\vspace*{-8mm}
\textcolor[rgb]{0.00,0.00,0.00}{\begin{center}
\includegraphics[width=.52\textwidth]{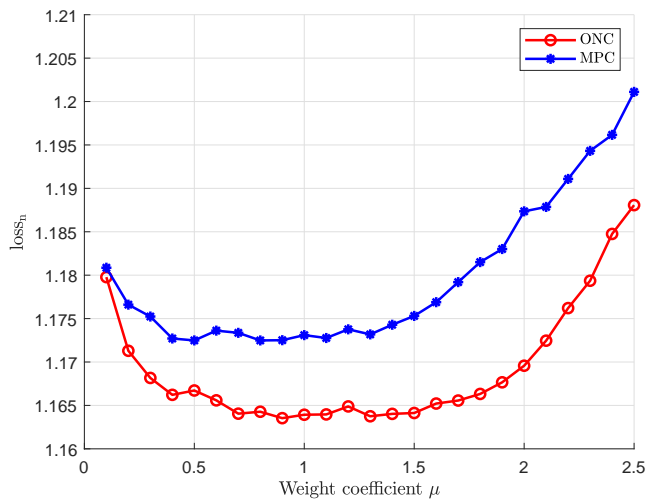}
\end{center}
\vspace{-4mm}
\caption{Narrow beam prediction loss as function of weight coefficient $\mu$ for the enhanced adaptive CBT scheme with $K = 7$.}
\label{fig7}}
\vspace{-4mm}
\end{figure}

\begin{table*}[tp!]\scriptsize 
\vspace{-1mm}
\centering
\renewcommand\arraystretch{1.0}
\makeatletter\def\@captype{table}\makeatother\caption{Complexity of proposed schemes.}\label{table3}
\vspace{-2mm}
\begin{tabular}{cccccc}
\toprule[0.8pt]
\rule{0pt}{11pt}
& \tabincell{c}{CNN assisted CBT\\of Sec. III}  & \tabincell{c}{LSTM assisted CBT\\of Sec. IV} & \tabincell{c}{Adaptive CBT\\of Sec. V} & \tabincell{c}{Enhanced adaptive CBT\\of Sec. V}              \\
\toprule[0.8pt]
Model size (MB) & 0.264 & 4.280 & 4.280 & 8.314               \\
Number of FLOPs (M) & 0.332 & 1.522 & 1.522 & 2.720                \\
Training time per epoch (s) & 8.855 & 13.536 & 15.870 & 26.591                  \\
Execution time per prediction ($\mu$s) & 24.749 & 34.315 & 34.315 & 47.680                   \\
\toprule[0.8pt]
\end{tabular}
\vspace{-4mm}
\end{table*}

The complexity of our proposed schemes are summarized in Table~\ref{table3} in terms of model size, number of floating-point operations (FLOPs), training time per epoch, and execution time per prediction measured on NVIDIA GeForce GTX 1080Ti. Note that the adaptive CBT scheme uses the same model structure as the LSTM assisted CBT scheme, but it requires extra training time for the selection of trained wide beams. It can be seen that the sizes of all the proposed deep learning models are smaller than $10$\,MB, and the training times per epoch are all less than $30$\,s, which indicates that our proposed deep learning models can be quickly trained and easily deployed. Besides, the execution times for each prediction are all smaller than $50$\,$\mu$s, which ensures fast beam alignment in mobile scenarios.

\subsection{Investigation of Our Adaptive CBT Schemes}\label{S6.3}

Four adaptive CBT schemes are actually proposed in Section~\ref{Sec:SCHEME3}, namely, the adaptive CBT with ONC, adaptive CBT with MPC, enhanced adaptive CBT with ONC and enhanced adaptive CBT with MPC. We now compare these adaptive CBT schemes.

\begin{figure}[tp!]
\vspace*{-1mm}
\begin{center}
\includegraphics[width=.5\textwidth]{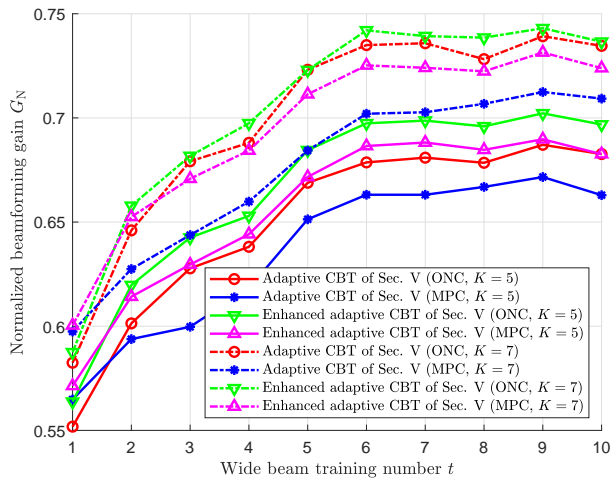}
\end{center}
\vspace{-4mm}
\caption{Normalized beamforming gain performance as function of wide beam training number for various adaptive CBT schemes given $K=5$ and $7$ trained wide beams. No additional narrow beam training, i.e, $K_\text{n}=0$.}
\label{Fig:simulation1} 
\vspace{-4mm}
\end{figure}

We first investigate the impact of the number of wide beam trainings $t$ on the normalized beamforming gain $G_{\text{N}}$ in Fig.~\ref{Fig:simulation1} for the four schemes, given the number of trained wide beams $K=5$ and $7$ as well as no additional narrow beam training. As expected, $G_{\text{N}}$ increases with $t$ for both the ONC and MPC based schemes, because more prior received signals provide more accurate UE movement information. The results thus demonstrate that both ONC and MPC can select the wide beams with high SNRs effectively. After $t=6$, $G_{\text{N}}$ appears to converge in all the cases. It can also be seen that increasing the number of trained wide beams from $K=5$ to $K=7$ improves the achievable $G_{\text{N}}$. Moreover, the performance of the enhanced adaptive scheme is better than its basic counterpart, which validates that auxiliary LSTM is capable of enhancing the accuracy of tracking UE location in mobile scenarios. Also the ONC based scheme outperforms its MPC based counterpart.

\begin{figure}[tp!]
\vspace*{-1mm}
\begin{center}
\includegraphics[width=.5\textwidth]{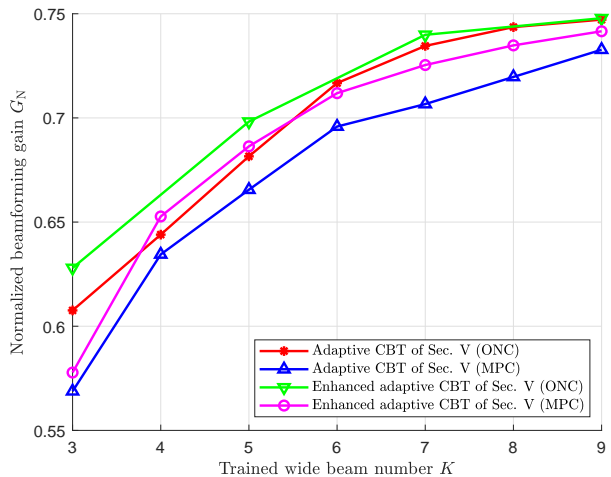}
\end{center}
\vspace{-4mm}
\caption{Normalized beamforming gain performance averaged over $t = 6$ to $10$ as function of trained wide beam number for various adaptive CBT schemes. No additional narrow beam training, i.e, $K_\text{n}=0$.}
\label{Fig:simulation2} 
\vspace{-4mm}
\end{figure}

\begin{figure}[tp!]
\vspace*{-1mm}
\begin{center}
\includegraphics[width=.5\textwidth]{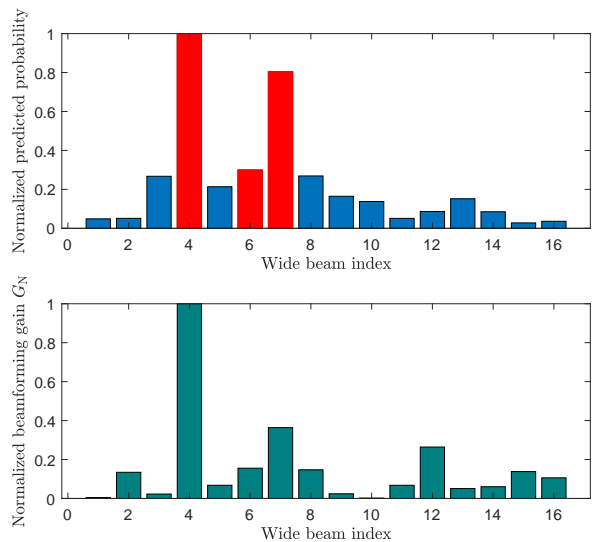}
\end{center}
\vspace{-4mm}
\caption{Comparison between normalized predicted probabilities and normalized beamforming gains of wide beams for the MPC based enhanced adaptive CBT scheme, where red columns denote the selected wide beams to be trained.}
\label{Fig:example} 
\vspace{-4mm}
\end{figure}

Fig.~\ref{Fig:simulation2} shows the impact of the number of trained wide beams $K$ on the achievable $G_{\text{N}}$, where the results are obtained by averaging $G_{\text{N}}$ over $t \geq 6$. Due to the symmetry of neighboring wide beams, only the results with odd $K$ are shown for the ONC based enhanced adaptive scheme. It can be seen that $G_{\text{N}}$ increases with $K$ since the deep learning model can extract more robust features from the received signals of more wide beams. The results confirm that the enhanced adaptive scheme outperforms its basic counterpart. Also the performance of the ONC based scheme is better than its MPC based counterpart, especially when $K$ is small. This is because noise and multipath interference may lead MPC to select irregular indices of wide beams, which may bring difficulties for the deep learning model to extract stable features. An example of $K=3$ is illustrated in Fig.~\ref{Fig:example}, where the predicted probabilities have several local maximums due to the noise and NLOS paths. This may make MPC ignore the neighboring wide beams of the strongest wide beam and lead the prediction model to fail to track the beam switch in mobile scenarios.

The results of this investigation suggest that among the four adaptive CBT schemes introduced in  Section~\ref{Sec:SCHEME3}, the ONC based enhanced adaptive scheme performs the best. Accordingly, we use this scheme to represent our adaptive CBT approach in the following performance comparison.

\begin{figure}[tp!]
\vspace*{-1mm}
\begin{center}
\includegraphics[width=.5\textwidth]{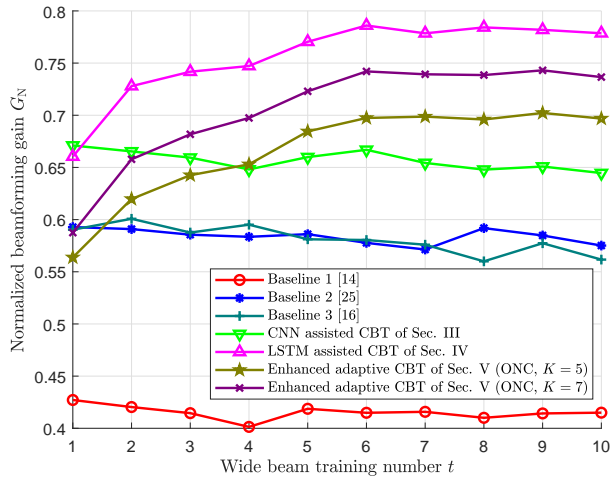}
\end{center}
\vspace{-4mm}
\textcolor[rgb]{0.00,0.00,0.00}{\caption{Normalized beamforming gain performance as function of wide beam training number for various schemes with $K_\text{n}=0$.}
\label{Fig:simulation3}} 
\vspace{-4mm}
\end{figure}

\subsection{Performance Comparison}\label{S6.4}

We compare the performance of our proposed schemes with the following three baselines:
\newline{Baseline 1: The noise-free beam prediction scheme in \cite{Ref:CALIBRATED_TRAINING} with $N_{\text{Tx}} / s_{\text{Tx}}$ measurements of wide beams.}
\newline{Baseline 2: The deep learning based beam prediction scheme in \cite{Ref:BEAM_PREDICTION7} with $N_{\text{Tx}} / s_{\text{Tx}}$ measurements of uniformly sampled narrow beams.}
\newline{Baseline 3: The adaptive and sequential beam alignment scheme in \cite{Ref:R4_2} with $N_{\text{Tx}} / s_{\text{Tx}}$ measurements of hierarchical beams and the target resolution $N_{\text{Tx}}$.}

With no additional narrow beam training, i.e, $K_\text{n}=0$, we first investigate the impact of the number of wide beam trainings $t$ on the achievable normalized beamforming gain $G_{\text{N}}$ in Fig.~\ref{Fig:simulation3}, where all the three baselines and our CNN assisted scheme do not rely on the prior information and thus their $G_{\text{N}}$ performance does not change with $t$. It can be seen that all our deep learning schemes significantly outperform all the three baselines. Specifically, the performance of the deep learning based baseline 2 and the adaptive alignment based baseline 3 are dramatically better than baseline 1, but $G_{\text{N}}$ of our CNN assisted scheme is 5\% higher than those of baseline 2 and baseline 3. Moreover, the LSTM assisted scheme attains the best performance, and the enhanced adaptive scheme has the second-best performance. Similar to the enhanced adaptive scheme, $G_{\text{N}}$ of the LSTM assisted scheme increases with $t$ and it converges after $t\ge 6$.

\begin{figure}[tp!]
\vspace*{-1mm}
\begin{center}
\includegraphics[width=.5\textwidth]{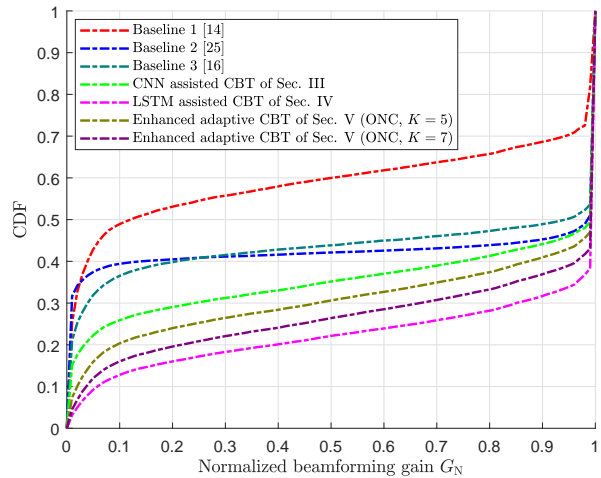}
\end{center}
\vspace{-4mm}
\textcolor[rgb]{0.00,0.00,0.00}{\caption{Comparison of the CDFs of the predicted narrow beam gains for various schemes with $K_\text{n}=0$.}
\label{Fig:simulation4}} 
\vspace{-4mm}
\end{figure}

According to Fig.~\ref{Fig:simulation3}, the normalized beamforming gains $G_{\text{N}}$ of all the models converge after $t \ge 6$. Therefore, in all the following simulation experiments, we use the results of $G_{\text{N}}$ after the models have converged, i.e., we use the average values of $G_{\text{N}}$ over $t=6$ to $10$.

\begin{figure}[tp!]
\vspace*{-1mm}
\begin{center}
\includegraphics[width=.5\textwidth]{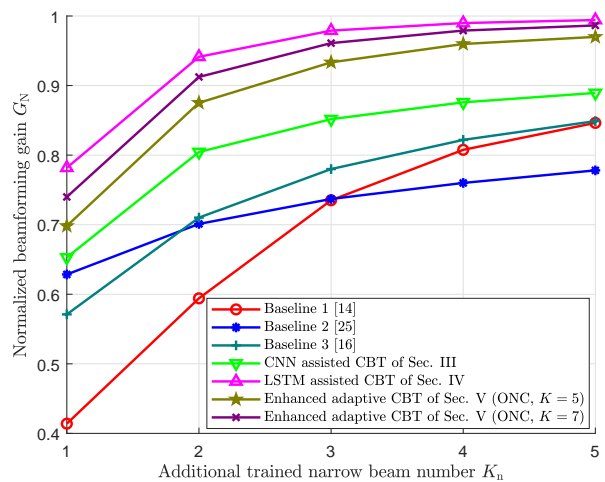}
\end{center}
\vspace{-4mm}
\textcolor[rgb]{0.00,0.00,0.00}{\caption{Normalized beamforming gain performance versus number of additional trained narrow beams for various schemes.}
\label{Fig:simulation5}} 
\vspace{-4mm}
\end{figure}

The superiority of our proposed schemes over the three baselines is further demonstrated in the cumulative distribution functions (CDFs) of the predicted narrow beam gains of various schemes shown in Fig.~\ref{Fig:simulation4}. The following general observations can be drawn from Figs.~\ref{Fig:simulation3} and \ref{Fig:simulation4}. Our CNN assisted scheme achieves better beam prediction than baseline 1, since it can extract more robust features from the received signals of all the wide beams. Our CNN assisted scheme can estimate the range of the optimal narrow beam more accurately than baseline 2 and it still obtains the suboptimal narrow beam even when the beam direction is not perfectly aligned. Thus our CNN assisted scheme outperforms baseline 2 because wide beams can achieve greater angular coverage than the sampled narrow beams used in baseline 2. Baseline 3 performs worse than our CNN assisted scheme due to its under-exploration of the beam space. The comparison between the CNN assisted scheme and the LSTM assisted scheme demonstrates that the prior information can effectively reduce incorrect predictions. It can also be seen that the enhanced adaptive scheme significantly reduces the training overhead at the cost of some degradation in prediction accuracy.

In our proposed schemes, the additional narrow beam training can be performed to further calibrate beam directions according to the predicted probabilities. Here, we investigate the impact of the number of additional trained narrow beams $K_\text{n}$ on the normalized beamforming gain $G_{\text{N}}$ in Fig.~\ref{Fig:simulation5}. As expected, the additional narrow beam training improves the achievable $G_{\text{N}}$ for all the schemes.
Observe that the performance gains of our deep learning based schemes over the existing deep learning based baseline 2 increase with $K_\text{n}$. Moreover, with $K_\text{n}=4$, the LSTM assisted scheme achieves almost the perfect beam alignment of $G_{\text{N}}=99.0\%$, and the enhanced adaptive scheme given $K=5$ attains $G_{\text{N}}=96.0\%$. The enhanced adaptive scheme of course has an additional advantage of requiring significantly lower overhead of beam training.

\begin{figure}[tp!]
\vspace*{-1mm}
\begin{center}
\includegraphics[width=.5\textwidth]{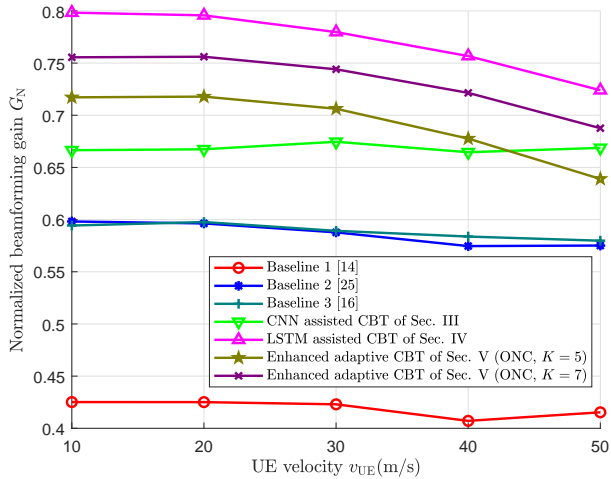}
\end{center}
\vspace{-4mm}
\textcolor[rgb]{0.00,0.00,0.00}{\caption{Normalized beamforming gain performance versus UE velocity for various schemes with $K_\text{n}=0$.}
\label{Fig:simulation6}} 
\vspace{-4mm}
\end{figure}

\begin{figure}[tp!]
\vspace*{-1mm}
\begin{center}
\includegraphics[width=.5\textwidth]{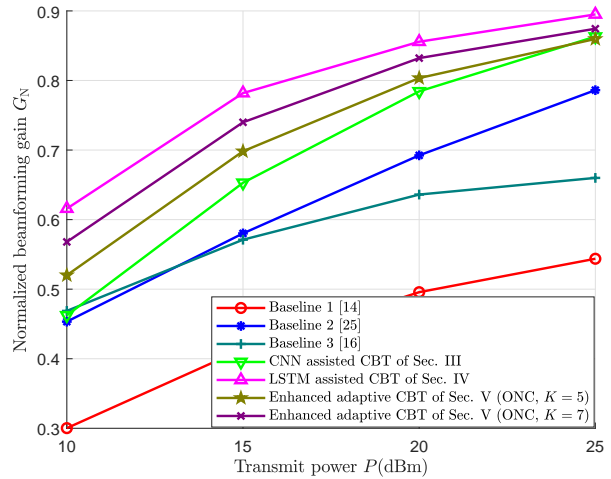}
\end{center}
\vspace{-4mm}
\textcolor[rgb]{0.00,0.00,0.00}{\caption{Normalized beamforming gain performance versus transmit power for various schemes with $K_\text{n}=0$.}
\label{Fig:simulation7}} 
\vspace{-4mm}
\end{figure}

Next, Fig.~\ref{Fig:simulation6} investigates the impact of UE velocity $v_\text{UE}$ on the achievable normalized beamforming gain $G_{\text{N}}$ performance by varying $v_\text{UE}$ from 10\,m/s to 50\,m/s. It can be seen that $v_\text{UE}$ does not affect the performance of the three baselines and our CNN assisted scheme much, since these schemes do not rely on the prior information. By contrast,  $v_\text{UE}$ has some adverse effect on our LSTM assisted scheme and enhanced adaptive scheme. Specifically, when $v_\text{UE}$ increases from 10\,m/s to 50\,m/s, their $G_{\text{N}}$ performance reduces by around 7\%. This is because that UE movement information is more difficult to accurately extract under high UE velocities.

We also investigate the impact of the transmit power $P$ on the normalized beamforming gain $G_\text{N}$ in Fig.~\ref{Fig:simulation7} by varying $P$ from 10\,dBm to 25\,dBm. Obviously, $G_{\text{N}}$ increases with $P$ owing to higher SNRs. It can be seen that our CNN assisted scheme achieves larger performance enhancement than the existing deep-learning based baseline 2 and adaptive alignment based baseline 3 as $P$ increases, which further verifies the advantage of our proposed scheme in high SNR scenarios over baseline 2 and baseline 3.

\begin{figure}[tp!]
\textcolor[rgb]{0.00,0.00,0.00}{\vspace*{-1mm}
\begin{center}
\includegraphics[width=.5\textwidth]{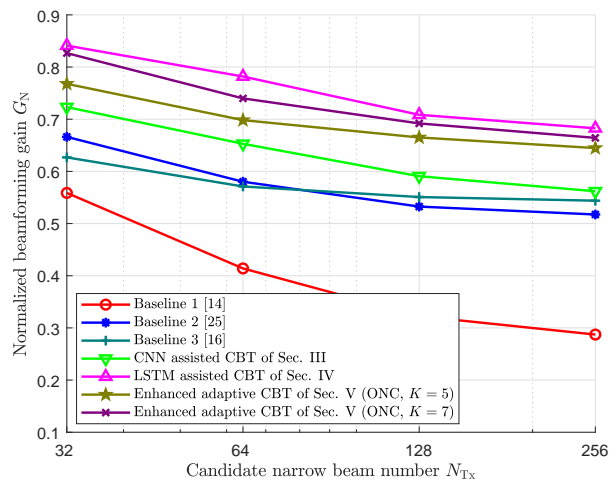}
\end{center}
\vspace{-4mm}
\caption{Normalized beamforming gain performance versus number of candidate narrow beams for various schemes with $K_\text{n}=0$.}
\label{fig16} 
\vspace{-4mm}}
\end{figure}

\begin{figure}[tp!] 
\textcolor[rgb]{0.00,0.00,0.00}{\vspace*{-1mm}
\begin{center}
\includegraphics[width=.5\textwidth]{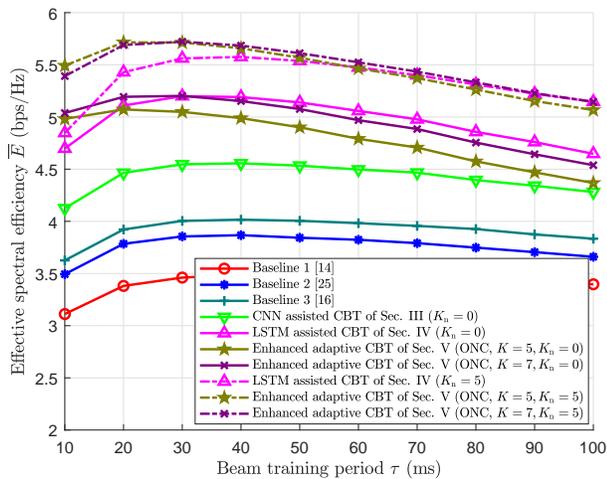}
\end{center}
\vspace{-4mm}
\caption{Effective spectral efficiency over $1,000$\,ms from initial access as function of beam training period for various schemes.}
\label{fig17}
\label{Fig:ASE}
\vspace{-4mm}}
\end{figure}

The achievable normalized beamforming gains $G_\text{N}$ for different candidate narrow beam numbers $N_{\text{Tx}}$ with $K_\text{n}\! =\! 0$ are depicted in Fig.~\ref{fig16}, where the wide beam number $N_{\text{Tx}} / s_{\text{Tx}}$ is fixed to $16$. Obviously, $G_\text{N}$ decreases as $N_{\text{Tx}}$ increases for all the schemes. Our schemes clearly outperform the three baselines for all $N_{\text{Tx}}$, which demonstrates the scalability of our prediction models.

Finally, we investigate the achievable effective spectral efficiency performance, $\overline{E}$ of (\ref{eq43}), for various schemes over $T_{\rm tot}\! =\! 1,000$\,ms from initial access. Specifically, Fig.~\ref{fig17} depicts the effective spectral efficiency performance as the functions of the beam training period $\tau$. It can be seen that $\overline{E}$ initially increases with $\tau$ owing to reduced beam training overhead. However, after achieving the maximum $\overline{E}$, increasing $\tau$ further decreases $\overline{E}$. This is because the loss of beam alignment is more likely to occur during longer data transmission. Observe that our enhanced adaptive CBT scheme with additional narrow beam training achieves the largest $\overline{E}$ especially under small $\tau$, since it can achieve almost perfect beam alignment with smaller overhead of beam training.

\section{Conclusions}\label{Sec:CONCLUSION}

To reduce the overhead of mmWave beam training, a deep learning assisted calibrated beam training approach has been proposed in this paper, and the feasibility of estimating the angle of the dominant path based on the channel power leakage in the received signals of beam training has been elaborated. Three schemes have been designed to predict the optimal narrow beam according to the received signals of wide beam training, by utilizing deep learning models to handle the highly nonlinear properties of the channel power leakage. Specifically, CNN has been adopted in the first scheme to predict the beam based on the instantaneous received signals. Furthermore, the additional narrow beam training according to the predicted probabilities has been proposed to further calibrate beam directions. In the second scheme, LSTM has been adopted to track the movement of UE and calibrate the beam direction based on the received signals of prior beam training. In the third scheme, an adaptive beam training strategy has been proposed where partial wide beams are selected to be trained based on the prior received signals. Two criteria, namely, ONC and MPC, have been designed for the selection, where ONC selects the neighboring wide beams of the predicted optimal beam, while MPC selects the wide beams with the top predicted probabilities. To better cope with UE mobility, auxiliary LSTM has been introduced to calibrate the directions of the selected wide beams more precisely. Simulation results have demonstrated that our proposed deep-learning based schemes achieve significantly higher beamforming gain while imposing smaller beam training overhead, compared with the conventional and existing deep-learning based beam training schemes.

{

}

\end{document}